\documentclass[a4paper, 11pt]{article}

\usepackage{preamble}

\title{Polarity and anti-distortive polarons in \ch{WO3} through epitaxial shear strain}

\author[1,6,7]   {Ewout van der Veer        \orcidlink{0000-0001-6634-113X} \textsuperscript{*} \footnote{ewout.vanderveer@uni-due.de} }
\author[1,2] {Martin F. Sarott              \orcidlink{0000-0001-9660-6346} \footnote{m.f.sarott@rug.nl} }
\author[3,5] {Jack T. Eckstein              \orcidlink{0000-0002-4814-4324} \footnote{ecksteinjt@ornl.gov} }
\author[1]   {Stijn Feringa                 \orcidlink{0009-0007-5605-6838} \footnote{s.feringa@rug.nl} }
\author[1]   {Dennis van der Veen                                           \footnote{d.van.der.veen.8@student.rug.nl} }
\author[1]   {Johanna van Gent González     \orcidlink{0000-0002-1873-1560} \footnote{j.van.gent.gonzalez@rug.nl} }
\author[1,2] {Majid Ahmadi                  \orcidlink{0000-0003-2321-3060} \footnote{majid.ahmadi@rug.nl} }
\author[1,2] {Horatio R. J. Cox             \orcidlink{0000-0002-1257-6299} \footnote{h.r.j.cox@rug.nl} }
\author[4]   {Ellen M. Kiens                \orcidlink{0000-0003-3106-6494} \footnote{e.m.kiens@utwente.nl} }
\author[4]   {Gertjan Koster                \orcidlink{0000-0001-5478-7329} \footnote{g.koster@utwente.nl}}
\author[1,2] {Bart J. Kooi                  \orcidlink{0000-0002-0311-4105} \footnote{b.j.kooi@rug.nl} }
\author[3]   {Michael A. Carpenter                                          \footnote{mc43@esc.cam.ac.uk} }
\author[3]   {Ekhard K. H. Salje†           \orcidlink{0000-0002-8781-6154} }
\author[1,2] {Beatriz Noheda                \orcidlink{0000-0001-8456-2286} \footnote{b.noheda@rug.nl} }

\affil[1]{Zernike Institute for Advanced Materials, University of Groningen, Groningen, The Netherlands}
\affil[2]{Groningen Cognitive Systems and Materials Center (CogniGron), University of Groningen, Groningen, The Netherlands}
\affil[3]{Department of Earth Sciences, University of Cambridge, Cambridge, United Kingdom}
\affil[4]{MESA+ Institute for Nanotechnology, University of Twente, Enschede, The Netherlands}
\affil[5]{Center for Nanophase Materials Sciences, Oak Ridge National Laboratory, Oak Ridge, Tennessee, USA}
\affil[6]{Faculty of Physics, University of Duisburg-Essen, Duisburg, Germany}
\affil[7]{Research Center Future Energy Materials and Systems (RC FEMS), Research Alliance Ruhr, Bochum, Germany}

\date{}

\begin{document}

\maketitle

\newpage

\begin{abstract}
    Bestowing CMOS-compatible binary oxides with additional functionalities is a powerful strategy toward the realization of  oxide electronics.
    Ideal candidates are thin films which display a strong sensitivity to strain, chemical doping or nanoscale confinement.
    Among these, crystalline tungsten trioxide \ch{WO3} exhibits exceptional structural flexibility, enabling a wide range of functionalities.
    Here, we reveal the emergence of a previously unreported polar phase in epitaxial \ch{WO3} thin films.
    We accomplish this by imposing epitaxial shear strain, which stabilizes a low-symmetry triclinic structure that persists up to large film thicknesses and elevated temperatures.
    At the atomic scale, a change in the oxygen octahedral tilt pattern facilitates this symmetry lowering into a polar phase, which manifests as a periodic in-plane polarized stripe domain configuration with needle-like bifurcations at the microscale.
    The stripe domain walls further exhibit a strongly enhanced electrical conductivity in conjunction with a pronounced reduction of a distortive structural mode, providing the first experimental evidence for the formation of anti-distortive polarons recently predicted in \ch{WO3}.
\end{abstract}

\newpage

\section{Introduction}
Transition metal oxides exhibit a wide range of physical properties, making them attractive candidates for new electronic materials.
Real-world applications of this class of materials have, however, long been limited by their difficult integration with existing complementary metal-oxide semiconductor (CMOS) fabrication processes.
Recent efforts have demonstrated additional useful functionalities in simpler binary transition metal oxides that are innately CMOS-compatible via chemical doping, interface engineering, nanoscale confinement, or strain engineering strategies.
Nanoscale ferroelectricity in hafnia\autocite{boescke2011} and the metal-to-insulator transition in vanadium dioxide\autocite{maher2024} are the most prominent examples.
Another similar binary oxide is tungsten trioxide (\ch{WO3}), which is conventionally used for its electrochromic\autocite{granqvist2014, deb2008} and gas sensing\autocite{dong2020} properties, as well as increasingly in resistive random access memory (ReRAM) devices\autocite{ji2016, begon-lours2022}, though predominantly in amorphous or polycrystalline form.

In single crystal form, \ch{WO3} exhibits a large number of structural polymorphs, as a result of its relatively simple structure consisting of corner-sharing \ch{WO6} octahedra. The structure of \ch{WO3} can be thought of as a perovskite structure (\ie \ch{ABO3}) with a vacant A-site, commonly known as the \ch{ReO3} structure.
Within this general framework, there exist various different phases, distinguished by their respective patterns of octahedral tilts and distortions.
These phases and the phase transitions between them, have been extensively reviewed in literature\autocite{salje1975, diehl1978, woodward1995, locherer1999, howard2002, eckstein2022, hamdi2016}, with each of them being ferroelastic, leading to complex twinning patterns.
Of the known phases, only the lowest-temperature \textepsilon\ phase ($\textrm{T < }\qty{240}{\kelvin}$, space group \emph{Pc}) has been reported to be (weakly) polar\autocite{matthias1949, woodward1997}, though other accounts have suggested that the ground state phase has a centrosymmetric structure with space group \emph{P2\textsubscript{1}/n}.\autocite{eckstein2022, hamdi2016, eckstein2024, hassani2022}
Furthermore, the boundaries between ferroelastic domains in \ch{WO3} have shown intriguing functionalities not present in bulk, including locally enhanced conductivity\autocite{kim2010}, superconductivity\autocite{aird1998a, aird1998b, aird2000, mascello2020} and a recently discovered electrical polarization, induced by local strain gradients at the twin walls through an effect termed \emph{flexopiezoelectricity}.\autocite{yun2015,yun2020, seo2023, eckstein2024}

Despite having an extraordinary structural flexibility and remarkable electronic properties, \ch{WO3} has, thus far, mostly been studied in bulk, single-crystal samples, rather than in crystalline thin films, which are more relevant for applications in electronic devices.
Moreover, using epitaxial strain engineering to stabilize metastable phases, manipulate the domain configuration, or act on the domain-wall density in \ch{WO3} tends to be hindered by the sparsity of lattice-matching substrates.
Note that the room-temperature stable phase in bulk \ch{WO3} displays a \emph{P2\textsubscript{1}/n} monoclinic symmetry with \emph{a} = \qty{7.306}{\angstrom}, \emph{b} = \qty{7.522}{\angstrom}, \emph{c} = \qty{7.678}{\angstrom}, and \emph{\textbeta} = \qty{90.88}{\degree}, resulting in pseudo-cubic lattice parameters of $a_{pc} = \qty{3.653}{\angstrom}$, $b_{pc} = \qty{3.761}{\angstrom}$ and $c_{pc} = \qty{3.839}{\angstrom}$.
Consequently, for thin films of \ch{WO3} grown on most standard oxide substrates, even on those with small lattice parameters, such as sapphire\autocite{tagtstrom1999, garg2000}, \ch{SrTiO3}\autocite{garg2000, du2014, yang2017, kalhori2016, sun2024}, LSAT\autocite{leng2015, yang2017}, or \ch{LaAlO3}\autocite{li2015, yang2017, sun2024}, the lattice mismatch is too large for epitaxial growth to be achieved and the \ch{WO3} films grow unstrained and contain a large amount of structural defects.
Exceptions are (110)-oriented \ch{YAlO3} (YAO) substrates.
Coherently strained films have been successfully grown on these substrates \autocite{leng2015, yun2015, yun2020, sun2024}.
However, in this case, no deviations from the bulk stable monoclinic \emph{P2\textsubscript{1}/n} phase have been observed.

In this work, we show that it is possible to stabilize a new polar and piezoelectric phase in epitaxial \ch{WO3} thin films at room temperature.
This polar phase is the result of a symmetry lowering from monoclinic to triclinic, driven by epitaxial shear strain imposed when growing \ch{WO3} on (001)-oriented \ch{YAO} substrates.
Using a combination of scanning transmission electron microscopy (STEM), angle-resolved piezoresponse force microscopy (PFM), and large-scale X-ray diffraction (XRD) reciprocal space mapping, we determine the structure of this new polar phase from the atomic scale to the macroscale.
We find that the \ch{WO3} films consist of four equivalent triclinic domain variants that give rise to a stripe-like piezoelectric domain structure with an in-plane orientation of the polarization.
At the atomic level, the triclinic polar phase is characterized by a unique octahedral tilt pattern that gets progressively suppressed in the vicinity of the domain walls.
Furthermore, conductive atomic force microscopy (cAFM) and scanning electron microscopy (SEM) reveal an enhanced electrical conductivity at these domain walls despite the absence of a polar discontinuity at these domain walls.
Together with the marked reduction of the distortive structural mode at the domain walls unveiled by STEM, our electrical characterization provides the first experimental evidence for the existence of anti-distortive polarons, recently predicted by Hassani et al.\autocite{hassani2025}

\section{Results \& Discussion}
\subsection{Polar domain structure}
\begin{figure}
    \centering
    \includegraphics[width=0.9\textwidth]{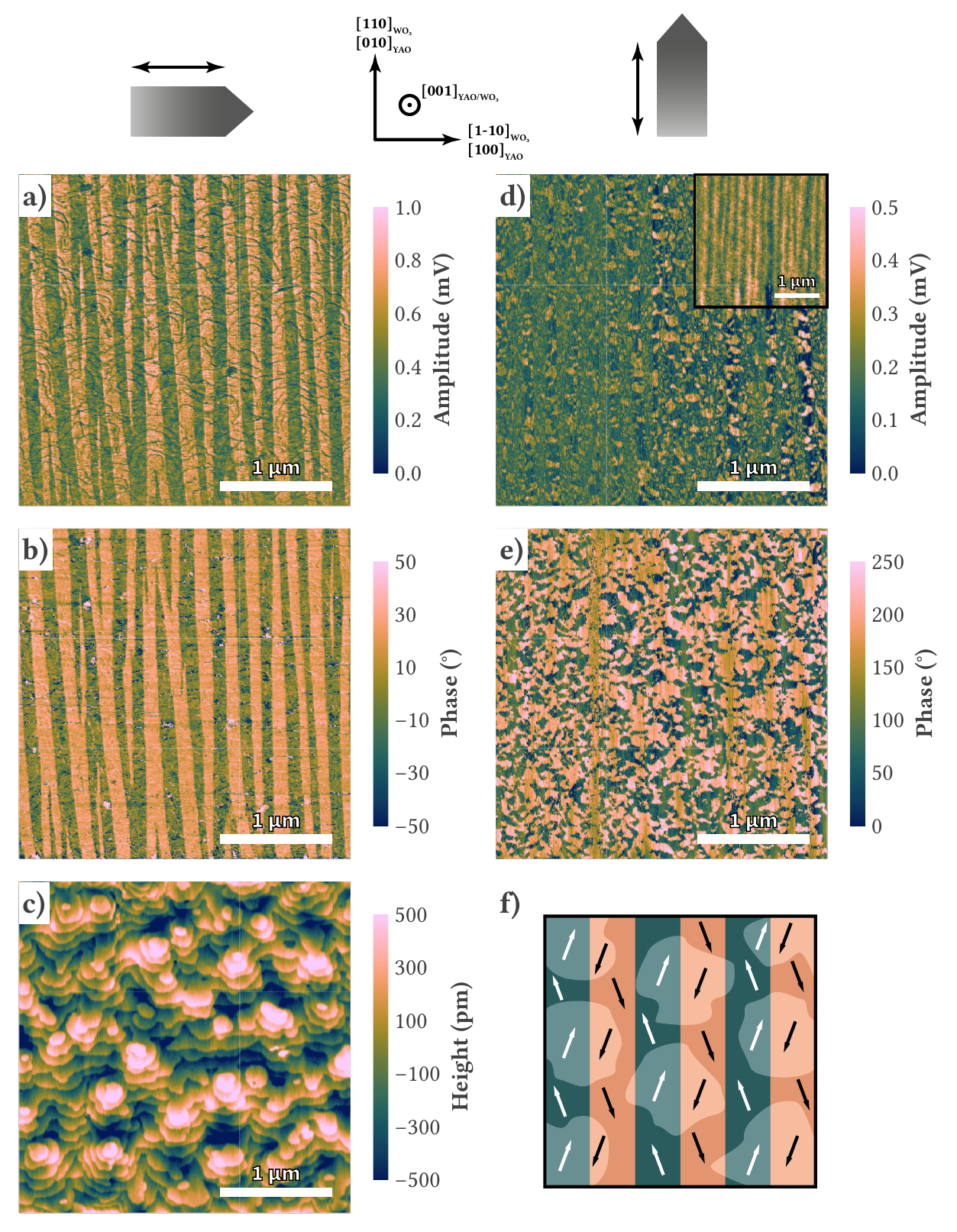}
    \caption{Angle-resolved lateral piezoresponse force microscopy (PFM) of a \ch{WO3} film on a (001)-oriented \ch{YAlO3} substrate at sample rotations of (a,b) \ang{0} and (d,e) \ang{90} with respect to the reference axes displayed on the top, showing lateral PFM (a,d) amplitude and (b,d) phase. The inset in (d) shows a vertical PFM amplitude image with a stripe-like buckling contrast corresponding to the domain configuration seen in (a), obtained via a \qty{90}{\degree} rotation of the fast scanning axis. c) Sample topography. f) Schematic illustration of the polarization vector. The polarization is predominantly oriented along the stripe domains seen in (a,b) with a small component in the orthogonal direction (d,e). }
    \label{fig:pfm}
\end{figure}

We begin our investigation of the \ch{WO3} films, grown via pulsed laser deposition on (001)-oriented \ch{YAO} (see Methods), by performing atomic force microscopy (AFM) and angle-dependent piezoresponse force microscopy (PFM), as shown in Fig.\ \ref{fig:pfm}.
For a film with a thickness of \qty{157}{\nano\meter}, the AFM topography (Fig.\ \ref{fig:pfm}c) shows an atomically flat surface with a wedding-cake-like morphology that is characteristic of two-dimensional layer-by-layer growth and in accordance with reflection high-energy electron diffraction measurements acquired in-situ during thin-film growth (see Fig.\ \ref{fig:suppl:rheed}).
The lateral PFM phase and amplitude images, taken with the cantilever axis oriented along the [100]\textsubscript{O} direction of the orthorhombic \ch{YAO} substrate (see Fig.\ \ref{fig:pfm}a,b), reveal clear domain contrast with long stripe-like domains and domain walls along the [010]\textsubscript{O} direction of the \ch{YAO}.
Since lateral PFM is sensitive to a projected polarization perpendicular to the cantilever axis, such a domain contrast indicates the presence of an in-plane polarization that lies along the stripe-domains and alternates in sign between neighboring domains.
To deepen our understanding of this domain configuration, we rotate the sample with respect to the cantilever axis and perform angle-dependent lateral PFM, as shown in Figs.\ \ref{fig:pfm}d,e and \ref{fig:suppl:pfm0deg}-\ref{fig:suppl:pfm90deg}.
Upon rotating the sample by \ang{90}, we detect additional mosaic-like domain features with an in-plane polarization component that lies perpendicular to that of the stripe domains along [100]\textsubscript{O} of \ch{YAO}.
Note that when changing the fast scanning axis to scan across the stripe domains for the film rotated by \ang{90}, the stripe domains reappear in the vertical channel as a buckling contrast (see inset of Fig.\ \ref{fig:pfm}d), which is consistent with an in-plane polarization as previously reported in other materials.\autocite{Nath2010, Gradauskaite2023}
Hence, as illustrated in Fig.\ \ref{fig:pfm}f, our lateral PFM data reveal the presence of a periodic stripe domain pattern in the \ch{WO3} films with an in-plane polarization that predominately lies along [010]\textsubscript{YAO} and exhibits a small component along [100]\textsubscript{YAO}, with irregular, meandering domain walls.
No domain contrast is found in vertical PFM (see Fig.\ \ref{fig:suppl:vertpfm}), but a small enhancement of the out-of-plane piezoresponse is present at the domain walls.
This local response is likely related to strain gradient-induced flexopiezoelectricity, as proposed by Yun \etal\autocite{yun2020} and observed in \ch{WO3} films on (110)-oriented \ch{YAO}.

Observing such a highly anisotropic domain pattern is rather surprising, given that the pseudo-cubic (001) cut of the \ch{YAO} substrate does not have an in-plane anisotropy along the pseudocubic unit cell axes.
Instead, the domain anisotropy likely originates entirely from the non-orthogonal base angle (\ang{91.5}) of the substrate and the resulting epitaxial shear strain imposed on the \ch{WO3} film.
Indeed, no comparable anisotropic domain pattern is observed in \ch{WO3} films on (110)-oriented \ch{YAO}\autocite{yun2020, yun2015}, where the two in-plane lattice constants are different ($a_{pc} = \qty{3.68}{\angstrom}$ vs. $b_{pc} = \qty{3.71}{\angstrom}$), but the base angle is \ang{90} and, thus, no shear strain is imposed.
The strong dependence of the polar domain structure on the \ch{YAO} base angle suggests that a shear strain-induced distortion of the \ch{WO3} unit cell is the primary driving force for the anisotropic domain formation.

The domain pattern further exhibits needle-like bifurcations, a feature commonly found in uniaxial polar and ferroelectric materials.\autocite{matsumoto2010, mcgilly2017,tikhonov2022}
For instance, the stripe-like domain pattern found here is reminiscent of that found in thin samples of \ch{BaTiO3} prepared by focused ion beam milling.\autocite{matsumoto2010}
In this system, the domain walls are vertical and along the in-plane polarization direction, so the domain walls maintain zero net charge.
New domains nucleate at sharp bifurcation points within domains of opposite polarity to minimize the surface area of the resulting charged domain wall, thereby minimizing the energy cost associated with the resulting uncompensated bound charge.
Note that tip-induced manipulation of the domain configuration using a DC electric field has been unsuccessful due to the relatively high conductivity of the \ch{WO3} film and the absence of a suitable bottom electrode.

\subsection{Macroscopic structural characterization}
\begin{figure}
    \centering
    \includegraphics[width=\textwidth]{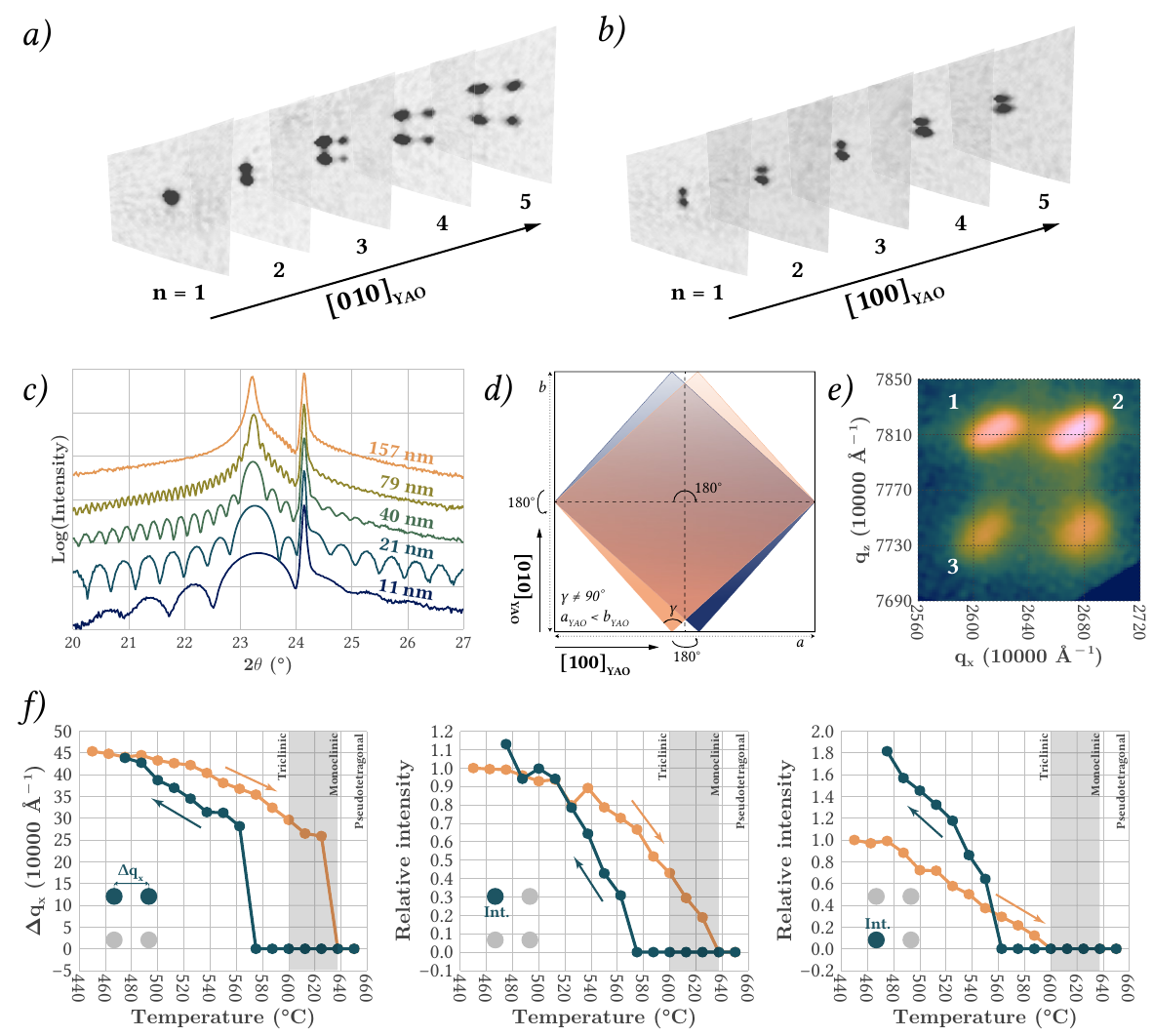}
    \caption{Structural characterization of \ch{WO3} films on (001)-oriented \ch{YAO}. a,b) Projections of reciprocal lattice planes of increasing order of a \qty{157}{\nano\meter} \ch{WO3} film along the a) [010]\textsubscript{YAO} and the [100]\textsubscript{YAO} directions. Each image represents the a) (2+n \textoverline{1}+n 4) or b) (\textoverline{1}+n 2-n 4) \ch{WO3} reflections, using pseudo-orthorhombic indices. The vertical and horizontal directions in each image are a) [001]\textsubscript{YAO} and [100]\textsubscript{YAO}; and b) [001]\textsubscript{YAO} and [010]\textsubscript{YAO}, respectively. c) Symmetric \straighttheta-2\straighttheta\ XRD line scans of the (002) reflection of \ch{WO3} films with varying thicknesses. d) Diagram showing the relationship between two structural \ch{WO3} domains and the \ch{YAO} substrate. The other two domains are related to the ones shown here by a \ang{180} rotation about the [001] axis perpendicular to the page. e) Reciprocal space map of the pseudo-cubic {103} reflections of the \ch{WO3} film at \qty{450}{\celsius} showing the four triclinic domains. The spots labeled '1', '2', and '3' were used to define the order parameters in (f). f) Evolution of the order parameters of the phase transition from the low-temperature triclinic phase into the high-temperature pseudo-orthorhombic phase.}
    \label{fig:structural}
\end{figure}

In bulk crystals of \ch{WO3} \autocite{eckstein2022}, a monoclinic phase with space group \emph{P2\textsubscript{1}/n} is stable at room temperature.
This phase is centrosymmetric and non-piezoelectric.
The formation of in-plane polarized domains with a distinctive piezoresponse (Fig.\ \ref{fig:pfm}) is, therefore, rather unexpected.
To shed light on the structural origins of this polar phase, we perform large scale reciprocal space mapping using a single-crystal X-ray diffractometer (see Methods) on a \qty{157}{\nano\meter} thick \ch{WO3} film (see Fig.\ \ref{fig:structural}).
Fig.\ \ref{fig:structural}a shows reconstructed images of reciprocal lattice planes spanned by the [001] (vertical) and [100] (horizontal) crystallographic directions of the \ch{YAO} unit cell at different offsets along the [010] direction, corresponding to the \nth{1} through the \nth{5} order of the same diffraction peak.
The \ch{WO3} peak clearly splits into four equivalent peaks in a square arrangement (\textbigsquare) with increasing distance along [010]\textsubscript{YAO}.
Fig.\ \ref{fig:structural}b shows the equivalent planes along the [100]\textsubscript{YAO} direction (\ang{90} rotated around the [001] axis normal to the sample surface), where a two-fold vertical splitting is seen (identical to the vertical splitting along the [010]\textsubscript{YAO} direction), but no change with increasing diffraction order.
This pattern is inconsistent with the monoclinic \emph{P2\textsubscript{1}/n} phase in bulk \ch{WO3}, where crystal twins would give rise to two-fold splitting along [001] or [100] (leading to four-fold splitting in a cross (\textbigplus) arrangement).

The observed peak splitting (the peak positions and the evolution of the splitting with the amplitude of the q-vector) can only be explained if the symmetry of the crystal lattice is lower than monoclinic, \ie triclinic.
We indexed the peaks using the CrysAlisPro software package into four equivalent triclinic domains, with the lattice parameters listed in Tab.\ \ref{tab:tab1}.

\begin{table}[ht]
    \centering
    \caption{Crystallographic data for \ch{WO3} film on (001)-oriented \ch{YAlO3}.}
    \label{tab:tab1}
    \begin{tabular}{l l}
        \toprule
        Temperature                      & \qty{300}{\kelvin}              \\
        Lattice system                   & Triclinic                       \\
        Wavelength                       & \qty{1.54056}{\angstrom}        \\
        \textphi\ range                  & \ang{0} -- \ang{360}            \\
        2\straighttheta\ range collected & \ang{0.46} -- \ang{140.54}      \\

        \midrule[0.5\lightrulewidth]
        \multicolumn{2}{l}{\textbf{Index limits}}                          \\
        \emph{h}                         & -8 to 8                         \\
        \emph{k}                         & -8 to 8                         \\
        \emph{l}                         & 1 to 5                          \\

        \midrule[0.5\lightrulewidth]
        \multicolumn{2}{l}{\textbf{Lattice parameters}}                    \\
        a                                & \qty{7.323(6)}{\angstrom}       \\
        b                                & \qty{7.470(2)}{\angstrom}       \\
        c                                & \qty{7.682(2)}{\angstrom}       \\
        \textalpha                       & \qty{89.04(3)}{\degree}         \\
        \textbeta                        & \qty{89.16(3)}{\degree}         \\
        \textgamma                       & \qty{88.74(5)}{\degree}         \\
        V                                & \qty{419.8(4)}{\angstrom\cubed} \\
        \bottomrule
    \end{tabular}
\end{table}

A triclinic (space group \emph{P\textoverline{1}}) phase is known to exist in bulk \ch{WO3} between \qtyrange{230}{290}{\kelvin}, but the lattice parameters and unit cell volume of this reported phase are different from those measured here.\autocite{salje1975, diehl1978, woodward1995}
Also, the existence of a center of inversion in the reported \emph{P\textoverline{1}} phase precludes the presence of a piezoresponse and polar domains, as we observe it in Fig.\ \ref{fig:pfm}, which implies a \emph{P1} space group in our films.

Symmetric \straighttheta-2\straighttheta\ scans (Fig.\ \ref{fig:structural}c) show extended Laue oscillations over the full range of thicknesses, indicating that the films are of high quality and crystallinity.
Films with thicknesses ranging from \qtyrange[range-phrase=\ to\ ]{11}{157}{\nano\meter} show no detectable differences in the lattice parameter, or other signatures of strain relaxation, indicating that the same phase observed in the thickest film is present already in the thinner films (see also Fig.\ \ref{fig:suppl:rsm_thicknesses}).

Fig.\ \ref{fig:structural}d illustrates the epitaxial relation between the \ch{YAO} substrate (black square) and two of the triclinic domains of the \ch{WO3} film. The [100]\textsubscript{YAO} and [001]\textsubscript{YAO} axes are collinear with the [1\textoverline{1}0]\textsubscript{WO\textsubscript{3}} and [001]\textsubscript{WO\textsubscript{3}} axes, respectively. The [010]\textsubscript{YAO} axis is slightly offset from the [110]\textsubscript{WO\textsubscript{3}} axis in one of four directions, each corresponding to one of the four structural domains. The domains are related to each other by \ang{180} rotations about the three principal axes of the orthorhombic unit cell of the substrate.

We further characterize the temperature stability of the \emph{P1} phase by temperature-dependent X-ray reciprocal space mapping (RSM) around the (226) peak of the \ch{WO3} film (equivalent to the (103) peak of a pseudo-cubic perovskite).
In the room temperature triclinic phase, the RSM contains four diffraction spots in a square arrangement, again belonging to the four structural domains (see Figs.\ \ref{fig:structural}e and \ref{fig:suppl:rsms}).
Upon heating, we observe a phase transition around \qty{600}{\celsius}, above which only the top-right diffraction spot remains.
To describe this phase transition in more detail, we define three different order parameters: 1) The separation between the top two diffraction spots, 2) the normalized intensity of the top-left diffraction spot and 3) the normalized intensity of the bottom-left diffraction spot.
The latter two intensities are measured relative to the intensity of the top-right diffraction spot, which is maintained throughout the transition, to account for any overall decrease of intensity due to misalignment or sample degradation.
The evolution of these order parameters is shown in Fig.\ \ref{fig:structural}f.
We find that the triclinic phase is maintained up to a temperature of \qty{600}{\celsius}, after which the film undergoes a broad, two-step phase transition that is completed at \qty{650}{\celsius}.
The bottom two spots disappear in the first step around \qty{600}{\celsius} on heating (order parameter 3), resulting in either a domain configuration with two triclinic domains or a monoclinic intermediate phase.
The top spots merge at around \qty{640}{\celsius} (order parameters 1 and 2), leaving a phase with orthorhombic or tetragonal symmetry.
Upon cooling, the top spot splits at \qty{580}{\celsius}, after which the bottom spots reappear at \qty{560}{\celsius}.
The intensity of the top spot returns to the same level as before the transition, whereas that of the bottom spot increases by a factor of almost two, indicating that the population of the different structural domains in the sample changes upon cycling through the transition.

The series of phase transitions found in our films differs markedly from that reported in bulk \ch{WO3}\autocite{eckstein2022}, for which the monoclinic \emph{P2\textsubscript{1}/n} phase is stable between \qty{290}{\kelvin} and \qty{338}{\kelvin} and the only known triclinic \emph{P\textoverline{1}} phase exists between \qtyrange{230}{290}{\kelvin}.
The difference in the phase transition temperature seen here demonstrates that the epitaxial shear strain plays a fundamental role in the stability of the triclinic \emph{P1} phase in the \ch{WO3} films.

\subsection{Nanostructural characterization}
Having established symmetry lowering into a triclinic phase as the structural origin of the emergence of polar domains in Fig.\ \ref{fig:pfm}, we next move to scanning transmission electron microscopy (STEM) measurements to gain insights into the atomic-scale mechanism for strain accommodation.
\begin{figure}
    \centering
    \includegraphics[width=\textwidth]{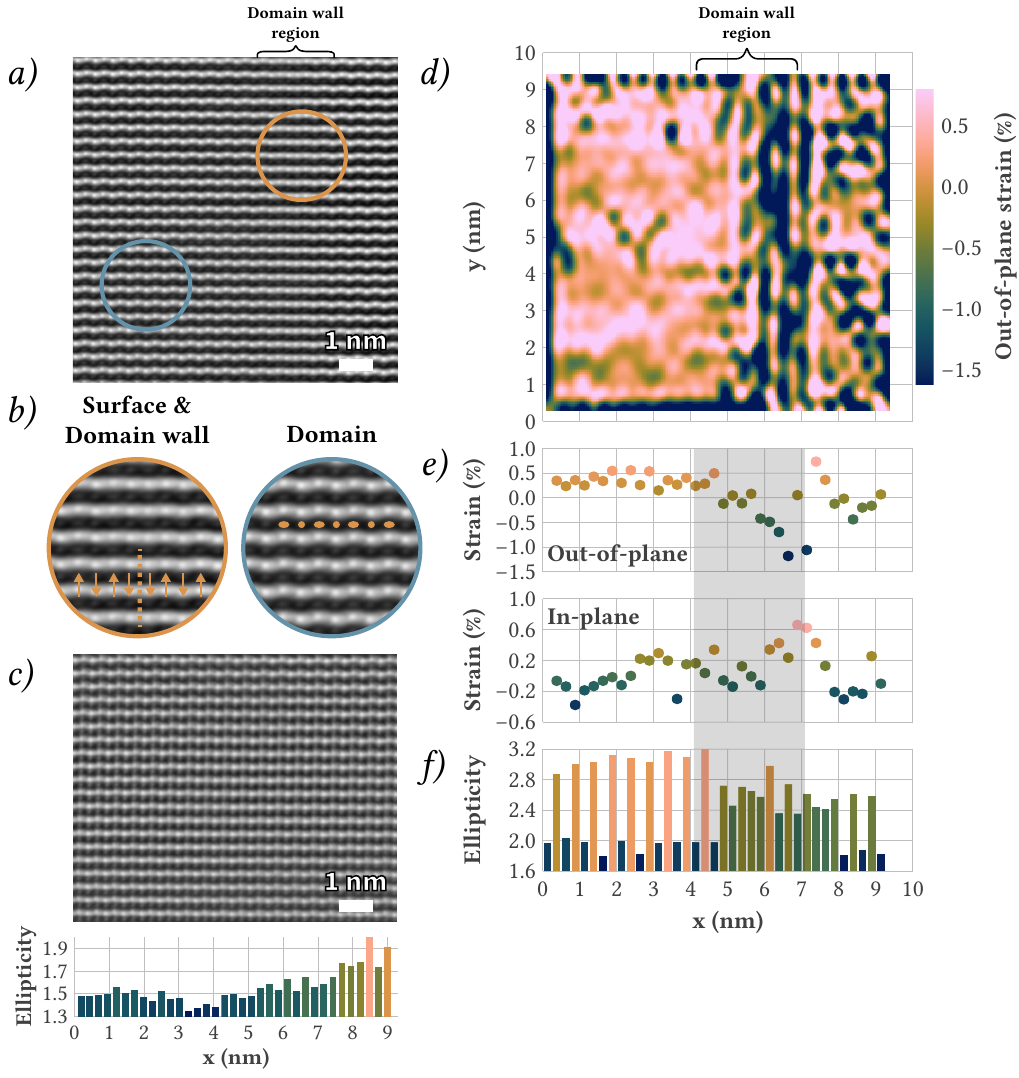}
    \caption{Atomic resolution structural characterization of the \ch{WO3} film on (001)-oriented \ch{YAO}. a) iDPC-STEM images of the film near a domain wall. b) Blow-ups of the image in (a) showing the change of structure in the surface and domain wall region and in the interior of a domain. Arrows in the left panel show the direction of the zigzag modulation of W atomic column positions. A half-unit-cell phase shift of this modulation occurs at the dashed line. c) iDPC-STEM image along the in-plane zone axis perpendicular to that in (a), showing no modulation of the oxygen column ellipticity (shown at the bottom of the image). The colors of the bars are correlated with the magnitude of the ellipticity d) Map of the out-of-plane (\ie vertical in the image) strain in the image in (a). Strain is defined as the local lattice parameter divided by the average lattice parameter in the image. e) Vertically-averaged profiles of the out-of-plane and in-plane strain in the image in (a) showing out-of-plane contraction and in-plane expansion of the unit cell. The symbol colors are correlated to the strain magnitude f) Vertically-averaged ellipticity of atomic columns of oxygen in the \vao layers showing a modulation of the ellipticity in the interior of the domains, which disappears in the domain wall region.}
    \label{fig:stem}
\end{figure}
Fig.\ \ref{fig:stem}a shows an integrated differential phase contrast (iDPC)-STEM image of the film along the [110] zone axis near a domain wall, showing sublayers of \ch{WO2} and \vao, where V\hspace{-0.04cm}\textsubscript{A} are the vacancies in the A-sites related to the perovskite structure .
The \ch{WO2} sublayers exhibit a zigzag-like modulation of W atomic positions that originates from the antipolar M3 distortion present in all reported low-symmetry phases of \ch{WO3} (see also Fig.\  S\ref{fig:suppl:bulkstructure}).\autocite{eckstein2022}
A progressive suppression of this zigzag modulation is visible in a vertical region (indicated in Fig.\ \ref{fig:stem}a as \emph{domain wall region}) of about eight pseudo-cubic unit cells in width.
This region is associated with a half-unit-cell phase shift of the modulation, \ie the region to the left of the domain wall is related to that on the right by a \ang{180} rotation about the [110]\textsubscript{\ch{WO3}} (zone axis) direction or the [1\textoverline{1}0]\textsubscript{\ch{WO3}} direction (horizontal in the image).
Such a \ang{180} rotation is exactly the relation between two structural twins found by X-ray diffraction (Fig.\ \ref{fig:structural}d), suggesting that this region of significantly reduced W-column modulation corresponds to a structural domain wall and, in particular, the domain wall between two polar stripe domains in Fig.\ \ref{fig:pfm}b.
Careful tracing of the atomic columns allows us to pinpoint the exact position where the distortion changes direction and shows that the actual twin wall is in fact a single unit cell thick (left panel in Fig.\ \ref{fig:stem}b).
A similar suppression of the zigzag modulation is also found in the first three to four unit cells near the surface of the film (Fig.\ \ref{fig:suppl:stem_surface}).

To further investigate the structural changes in the film near the domain wall, we perform real-space image analysis on the STEM image in Fig.\ \ref{fig:stem}a and extract the local lattice parameters (\ie strain) as a function of the position in the image, as well as the ellipticities of oxygen atomic columns in the \vao sublayers.
Fig.\ \ref{fig:stem}d shows a map of the out-of-plane strain across the domain wall.
The out-of-plane lattice parameter decreases in the vicinity of the domain wall while it is constant in the interior of the domain.
Integrating the strain map in the vertical direction yields the strain profile across the domain wall in Fig.\ \ref{fig:stem}e (top panel).
A profile of the in-plane strain made using the same procedure is shown in the bottom panel.
The contraction of the out-of-plane lattice parameter near the domain wall (gray shaded region in Fig.\ \ref{fig:stem}e,f) is associated with an expansion of the in-plane lattice parameter.

Furthermore, in the interior of the domains, we find that every second atomic column of oxygen in the \vao sublayers is broadened in the horizontal direction. That is, they exhibit an alternating broad/narrow pattern of ellipticity represented by broadened ellipses in Fig.\ \ref{fig:stem}b (right panel).
The magnitude of the ellipticity of each oxygen column in the \vao sublayer is measured by real-space image analysis and integrated over each vertical line to yield the bottom profile shown in Fig.\ \ref{fig:stem}f.
This alternating pattern of ellipticity---a direct consequence of the oxygen octahedral tilt pattern---is consistent with a triclinic \emph{P1} phase but inconsistent with the expected octahedral tilting patterns of both the monoclinic (\emph{P2\textsubscript{1}/n}) and bulk triclinic (\emph{P\textoverline{1}}) phases (see Fig.\ \ref{fig:suppl:bulkstructure})\autocite{glazer1972, woodward1997a}.
Moreover, as discernible in Fig.\ \ref{fig:stem}f, in the vicinity of the domain wall (gray band), this ellipticity alternation is markedly suppressed, along with the reduction of the zigzag motion of the W atomic columns.
Fig.\ \ref{fig:stem}(c) shows an iDPC-STEM image along the [1\textoverline{1}0] zone axis (\ie orthogonal to the STEM image in (a)) and corresponding oxygen column ellipticities.
In this zone, no comparable modulation of the ellipticities is present, demonstrating a clear structural anisotropy.
Our STEM data, thus, show an anisotropic polar structure with distinct higher-symmetry domain walls characterized by strongly reduced structural distortions.

\subsection{Domain width scaling with film thickness}
\begin{figure}
    \centering
    \includegraphics[width=\textwidth]{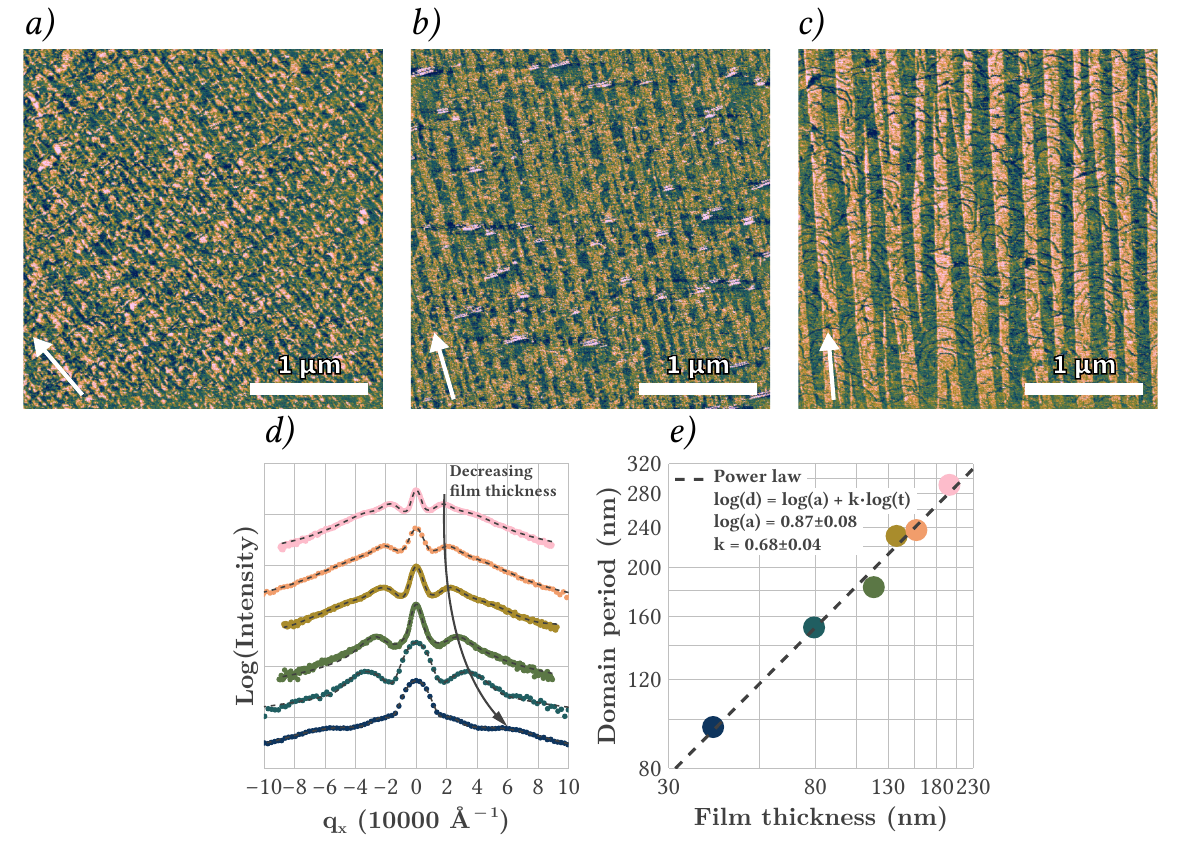}
    \caption{Domain width dependence on film thickness. (a-c) Lateral PFM amplitude images for a a) \qty{40}{\nano\meter}, b) \qty{79}{\nano\meter} and c) \qty{157}{\nano\meter} thick film. The arrows represent the predominant in-plane polarization axis. d) \textomega\ rocking curves around the (004) peak of the \ch{WO3} films, showing pronounced thickness-dependent domain satellite peaks. q\textsubscript{x} corresponds to the (1\textoverline{1}0) axis of the \ch{WO3} film. The dashed line represents the fit to the data. e) Log-log plot of the domain periodicity as a function of film thickness, extracted from the domain satellite peak positions in d). The dashed line is a fit to the data.}
    \label{fig:thickness}
\end{figure}

The dependence of domain size on film thickness can provide additional valuable insight into the driving force for the formation of a ferroelectric or ferroelastic domain structure and can demonstrate the ability for a material to be used in nanoscale electronics.
To investigate the domain size scaling in our \ch{WO3} films, we consider films with thicknesses ranging between \qtyrange[range-phrase=\ and\ ]{40}{196}{\nano\meter} and quantify the domain size in these films using PFM and X-ray \textomega\ rocking curves around the (004) peak of the \ch{WO3} film.
For large film thicknesses ($\mathrm{t_{film} \gg d_{DW}}$), the domain size in ferroelectrics and ferroelastics usually follows a power law with a scaling exponent of $\mathrm{\gamma = 0.5}$, in analogy with Kittel's law for ferromagnets.\autocite{catalan2007, schilling2006}
Figs.\ \ref{fig:thickness}a-c show lateral PFM images of films with different thicknesses, where the stripe domain width clearly increases with growing film thickness.
This is further confirmed by the X-ray \textomega\ rocking curve analysis, shown in Fig.\ \ref{fig:thickness}d.
Here, we observe the presence of pronounced satellite peaks in the [100]\textsubscript{YAO} direction with the peak position depending on the film thickness.
These peaks do not appear in the orthogonal [010]\textsubscript{YAO} direction and, therefore, must originate from the periodic stripe domains.
The domain period extracted from the spacing of the satellite peaks is, indeed, consistent with that observed in PFM (Fig.\ \ref{fig:thickness}e).
A power law fit to this data yields a scaling exponent of \qty{0.68 \pm 0.04}, similar to the value found for \ch{WO3} films on (110)-oriented \ch{YAO}.\autocite{yun2015}

Similar deviations from classical Kittel scaling ($\mathrm{\gamma = 0.5}$) have been seen in cases where domain walls are broadened due to magnetoelectric coupling\autocite{catalan2008} and in the thin-film (\ie $\mathrm{t_{film} \approx d_{DW}}$) regime due to ferroelastic effects and interactions between the substrate-film interface and the film surface.\autocite{pompe1993, pertsev1995, nesterov2013}
In our case, we do indeed also observe a widening of the domain walls (see Fig.\ \ref{fig:stem}a) that likely results from coupling between the structural and polar order parameters, giving rise to a strain gradient at the domain walls.
For thicknesses larger than \qty{250}{\nano\meter}, the film relaxes to the bulk monoclinic phase and no polar domains are present.

\subsection{Domain-wall conductivity and evidence for anti-distortive polarons}
\begin{figure}
    \centering
    \includegraphics[width=0.9\textwidth]{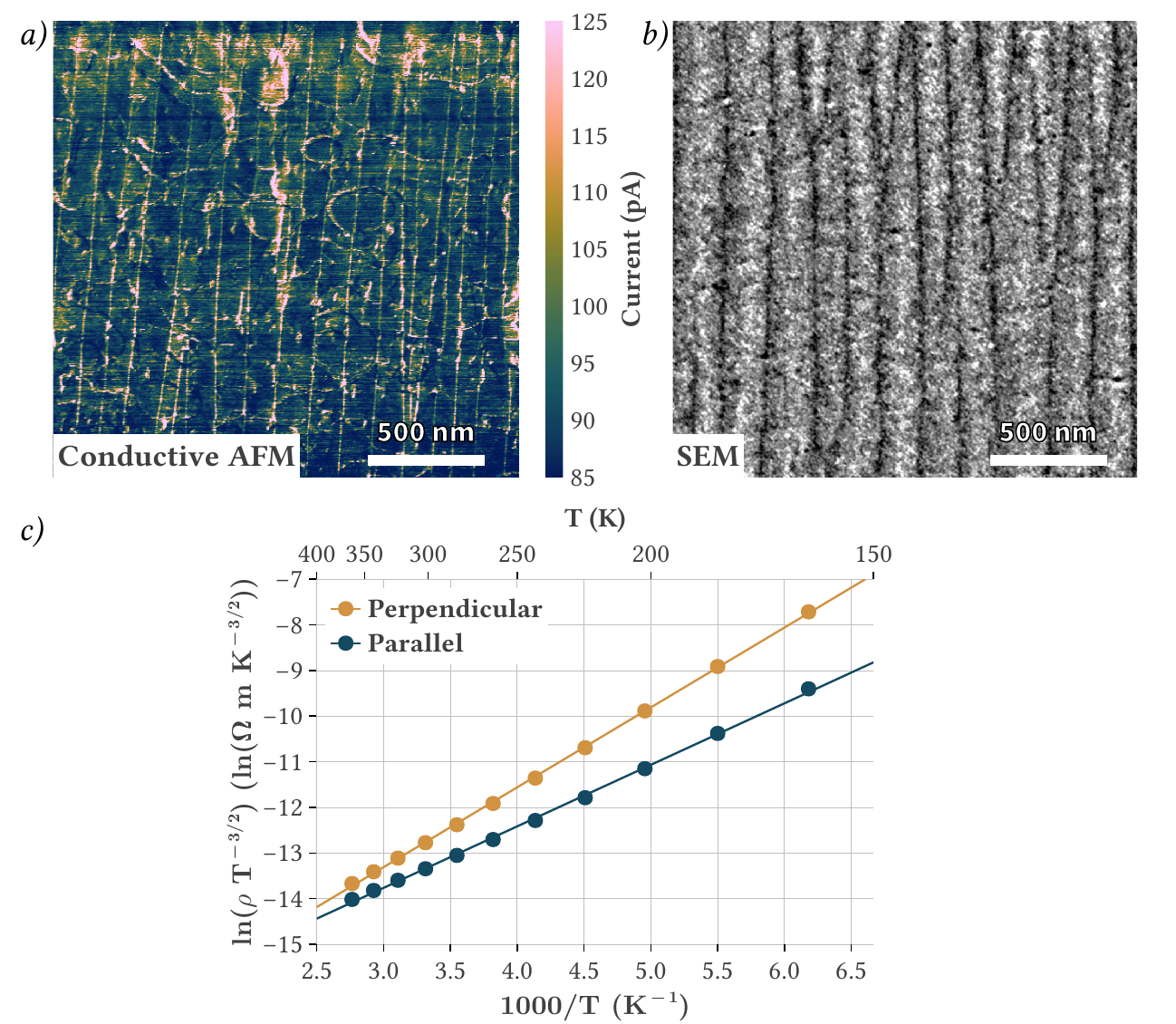}
    \caption{Micro- and macroscopic transport characteristics of the triclinic stripe domain walls in \ch{WO3}. a) cAFM image showing domain-wall contrast with an enhanced domain-wall conductivity. b) SEM image showing a reduced secondary electron yield at the location of the stripe domain walls. c) Temperature-dependent conductivity parallel and perpendicular to the stripe domain walls measured using Hall bars. The solid lines represent the fit according to a small polaron transport model.}
    \label{fig:conductivity}
\end{figure}

Domain walls in ferroic materials are characterized by a distinctly different symmetry than the interior of a domain, which can give rise to the emergence of additional functional properties.\autocite{Catalan2012,Nataf2020}
Moreover, domain walls in polar and ferroelectric materials frequently represent polar discontinuities that can repel or attract internal charge carriers or defects, and thus, exhibit locally modified conductance.\autocite{Seidel2009,Meier2012}
Even in ferroelastic materials, the conductivity of twin domain walls can be locally changed because of the accumulation of charged defects, such as oxygen vacancies.\autocite{kim2010}
To probe the local conducting properties of the stripe domain walls in our triclinic and polar \ch{WO3} films, we use conductive atomic force microscopy (cAFM) and scanning electron microscopy (SEM), as shown in Fig.\ \ref{fig:conductivity}a,b.
The cAFM image in Fig.\ \ref{fig:conductivity}a reveals an increased conductivity of the domain walls compared to the bulk of the domains, which appears further enhanced in the vicinity of bifurcations.
We also observe a minor increase of the detected current at topographical step edges, likely due to the locally increased tip-sample contact area.
The SEM image, shown in Fig.\ \ref{fig:conductivity}b, provides a complementary view of the local conductivity differences.
Here, we identify a clear decrease of the secondary electron yield at the domain walls.
Such a reduction is in good agreement with the expectation for a highly conductive region in which the primary electron beam is preferentially conducted away and, thus, the escape probability of secondary electrons is lowered.\autocite{vanRijn2024,Shih1997}
For in-plane ferroelectrics, the formation of charged head-to-head or tail-to-tail domain walls can lead to conducting domain walls and domain-wall contrast in SEM as previously reported.\autocite{hunnestad2020}
In our \ch{WO3} films, however, the domain walls are neutral without a polar discontinuity, which points to a different conduction mechanism at play.
Recent theoretical work by Hassani \etal\autocite{hassani2025} proposes the formation of anti-distortive polarons in \ch{WO3}, which are associated with a reduction of the structural distortions --- similar to those seen at the domain walls in our films (see Fig.\ \ref{fig:stem}).
Their work further suggests that this reduction of structural distortions is accompanied by a lowering of the conduction band minimum due to improved hybridization between O2p and W5d orbitals.

To gain insight into the conduction mechanism in our samples, we consider temperature-dependent electrical transport measurements parallel and perpendicular to the stripe domains using Hall bars, as shown in Fig.\ \ref{fig:conductivity}c and Fig.\ \ref{fig:suppl:hall}.
The temperature-dependent conductivity in Fig.\ \ref{fig:conductivity}c can be well fitted with a polaron transport model\autocite{Natanzon2020} (see Methods) for both orientations, whereas neither simple band gap activation nor variable range hopping can reproduce the data.
We further observe a distinctly reduced activation energy for transport parallel to the domain walls ($E_{a, \parallel} = \qty{0.1162 \pm 0.0012}{\electronvolt}$) in comparison to the direction perpendicular to the domain walls ($E_{a, \perp} = \qty{0.1508 \pm 0.0008}{\electronvolt}$) as well as a reduced charge carrier mobility although a polaronic conduction mechanism is at play in both cases.
This anisotropy can be attributed to the perpendicular path forcing charge carriers to hop from the conductive domain wall into the bulk of the domains, a process which constitutes a higher energy barrier than moving along the domain boundary network itself.
We suggest that the increased conductivity at domain walls in our film is due to a combination of band gap lowering associated with the reduction of octahedral tilting\autocite{hassani2025} and facilitation of anti-distortive polaron formation in the domain wall region.
Thus, given the clear indication for macroscopic polaronic transport in combination with enhanced domain wall conductivity and the undoing of the octahedral tilts at the domain walls, our work constitutes first experimental evidence for the formation of anti-distortive polarons.

\section{Conclusion}
We have demonstrated the stabilization of a novel polar phase in thin films of \ch{WO3} grown on (001)-oriented \ch{YAO}.
These films exhibit a stripe-like domain configuration with a near uniaxial in-plane polarization and needle-like bifurcations.
Structurally, this polar \ch{WO3} phase originates from a symmetry lowering into a triclinic system that is characterized by a unique pattern of oxygen octahedral tilts and gives rise to four structural domains.
Remarkably, we discover that the neutral \ch{WO3} stripe-domain walls exhibit enhanced conductivity, while simultaneously featuring a reduction of the distortive atomic modes inherent to the triclinic phase.
This combination, hence, points toward the first experimental evidence for the existence of anti-distortive polarons, as predicted theoretically.
These results constitute an important step towards the realization of domain wall-based nanoelectronics.

\section{Methods}
\paragraph{Thin film growth}
\ch{WO3} thin films were grown on the (001) face of as-received single-crystal substrates of \ch{YAO} (Crystec GmbH) by pulsed laser deposition using a \qty{248}{\nano\meter} \ch{KrF} excimer laser.
Films were grown from a ceramic \ch{WO3} target at a temperature of \qty{680}{\celsius}, an oxygen pressure of \qty{0.1}{\milli\bar}, a laser fluence of \qty{0.5}{\joule\per\centi\meter\squared} (spot size \qty{2.1}{\milli\meter\squared}), laser repetition rate of \qty{2}{\hertz} and a target-substrate distance of \qty{50}{\milli\meter}.
After growth, the films were kept at \qty{680}{\celsius} for up to \qty{30}{\minute}, then cooled to room temperature at a rate of \qty{10}{\celsius\per\minute} in growth pressure.
A total of between \qtyrange[range-phrase=\ and\ ]{200}{10000}{\pulses} were deposited, yielding final film thicknesses of \qtyrange{5}{196}{\nano\meter}.

\paragraph{X-ray diffraction}
Large-scale reciprocal space mapping was performed on a Bruker D8 Venture diffractometer equipped with a I\textmu S Cu microfocus X-ray source and a Photon II detector with sample placed at a \ang{20} angle of incidence to the X-ray beam.
Data are collected as two \ang{360} \textphi\ scans about an axis normal to the sample surface with a static detector placed at 2\straighttheta\ = \ang{36} and \ang{105}, similar to the method described by Sønsteby et al.\autocite{sonsteby2013}
2\straighttheta-\textomega\ line scans, reciprocal space maps and \textomega\ rocking curves were collected using a Panalytical X'Pert MRD thin-film diffractometer using a Cu source with a 2xGe(220) hybrid monochromator and a PIXcel\textsuperscript{3D} area detector in scanning line mode (2\straighttheta-\textomega\ line scans) or receiving slit mode (\textomega\ rocking curves). Fits were made to the rocking curves by least squares fitting of pseudo-Voigt functions.
Additional 2\straighttheta-\textomega\ scans and reciprocal space maps were measured with a Bruker D8 Discover diffractometer equipped with a rotating anode X-ray source (Cu-K\textalpha\ radiation), a Dectris Eiger2 R 500K two-dimensional area detector, a Ge(220) monochromator, and a \qty{1}{\milli\meter} beam collimator. For 2\straighttheta-\textomega\, the detector was used in 0D mode with a region of interest of 9x55 pixels and a fixed detector distance of \qty{295}{\milli\meter}. For reciprocal space maps, the detector was operated in 1D mode and kept stationary, while performing an \textomega\ rocking curve.

\paragraph{Electron microscopy}
Scanning electron microscopy images were collected on a FEI Helios G4 CX scanning electron microscope with an accellerating voltage of \qty{2}{\kilo\volt} and a beam current of \qty{1.4}{\nano\ampere}.
Cross-section ([110]\textsubscript{\ch{WO3}} zone) TEM lamellae were prepared by focussed ion beam milling (FEI Helios G4 CX) and Ga\textsuperscript{+} ion polishing at \qty{5}{\kilo\volt} and \qty{2}{\kilo\volt} to a thickness of approximately \qty{100}{\nano\meter}.
The TEM samples were plasma cleaned in \qty{25}{\percent}/\qty{75}{\percent} \ch{O2}/Ar atmosphere for \qty{1}{min} just before imaging.
A double-aberration-corrected ThermoFisher Themis Z scanning transmission electron microscope was operated at \qty{300}{\kilo\volt} with a beam current of \qty{50}{\pico\ampere} and a beam semiangle of \qty{24}{\milli\radian}.
High-angle annular dark field (HAADF) and integrated differential phase contrast\autocite{lazic2016} (iDPC) images were collected simultaneously at collection angles of \qtyrange{27}{164}{\milli\radian} and \qtyrange{7}{25}{\milli\radian}, respectively.
A four-segment annular dark field detector was used to collect iDPC images.
Samples were imaged in the [110]\textsubscript{\ch{WO3}} (cross-section) zone.

\paragraph{PFM and cAFM}
Scanning probe microscopy measurements were performed on an Asylum Research Cypher ES AFM using Bruker SCM-PIC-V2 \ch{PtIr}-coated conductive AFM probes.
PFM measurements were done in air and on-resonance with a tip-sample force of approximately \qty{15}{\nano\newton}.
Lateral and vertical PFM images were collected with the long axis of the cantilever along the [100], [010] and [110] in-plane directions of the substrate.
Additional measurements were performed using stiffer \ch{PtIr}-coated (Bruker SCM-PIT-V2) and \ch{CoCr}-coated (Bruker MESP-RC-V2) probes to exclude tip-specific artifacts which can arise when using soft probes (such as SCM-PIC-V2).
cAFM measurements were performed on the same instrument with SCM-PIC-V2 probes.
A small amount of silver paint was applied to part of the surface of the film, to which a positive electrical bias of \qtyrange{200}{800}{\milli\volt} was applied.
The current flowing from the sample into the grounded probe was measured while scanning.

\paragraph{Hall bar fabrication and electrical measurements}
Hall bar devices were fabricated on the \ch{WO3} films using a two-step electron-beam lithography (EBL) process, which defined channels with dimensions of \qty{150}\times \qty{30}{\micro\meter\squared}.
The procedure involved the e-beam evaporation and liftoff of Ti (\qty{5}{\nano\meter})/Au (\qty{40}{\nano\meter}) electrodes, followed by Ar ion-beam etching to isolate the active channel.
Temperature- and field-dependent transport properties were measured in a Quantum Design Physical Property Measurement System (PPMS) within a low-pressure ($<$\qty{4}{\milli\bar}) \ch{N2} environment from \qtyrange{100}{400}{\kelvin}.
Electrical contacts were made using ultrasonic Al wire bonding.
At each temperature, a Keithley 6221 Current Source and Agilent 3458A Voltmeter were used to perform four-probe measurements while sweeping the magnetic field from $-$\qty{9}{\tesla} to $+$\qty{9}{\tesla}.
Intrinsic electronic transport parameters were extracted using a two-stage linear regression analysis of the raw voltage data.
The process involved performing current sweeps at discrete magnetic fields, followed by a sweep of the field itself at each temperature.
To further improve accuracy and eliminate geometric errors, measurements were systematically performed across two sets of opposing voltage terminals.
This dual-sweep and multi-terminal methodology enables the systematic removal of extrinsic voltage offsets, including thermal and geometric artifacts, as detailed by Lindemuth.\autocite{lindemuth2020hall}
More details on the fitting procedure and the full analysis code are available on GitHub.\autocite{CoxGithub}

\paragraph{Data processing}
The CrysAlisPro software package was used to process single-crystal XRD data, to fit the lattice parameters of the four domains and to make simulated precession images from the data.
The STEM data were processed using ThermoFisher Velox software and in-house software.\autocite{vanderveerSTEMfit}
SEM data were processed using the Fiji\autocite{schindelin2012} software package.
The image was cropped and rotated, then denoised using Fourier filtering and Gaussian convolution.
Contrast in the image was enhanced using the \emph{contrast limited adaptive histogram equalization} (CLAHE) method.\autocite{zuiderveld1994}
PFM and cAFM data were processed using the Gwyddion\autocite{necas2012} software package.
The images were leveled by aligning rows and subtracting a second order polynomial from each.
Horizontal scan artifacts were removed.
The images were cropped and rotated where necessary to orient the stripe domains vertically.
All further processing and fitting of data was done using the Julia programming language\autocite{bezanson2017} using the GLM package\autocite{bates2023} for fitting.
Data were plotted using the Makie package\autocite{krumbiegel2021} with scientific color maps designed by Crameri.\autocite{crameri2018}

\section{Supporting information}
Supporting information are available from the Wiley Online Library or from the author.

\section{Acknowledgments}
Prof. Ekhard Salje passed away during the writing of this manuscript.
We dedicate this work to his memory.
The authors thank Jacob Baas, Henk Bonder and Joost Zoestbergen for technical support.
We thank Graeme Blake for helpful discussions.
E.vdV., M.F.S., M.A., H.R.J.C. and B.N. acknowledge fincancial support from the Groningen Cognitive Systems and Materials Center (CogniGron) and the Ubbo Emmius Foundation of the University of Groningen.
J.T.E., M.C., E.S. and B.N. acknowledge funding from the EU Horizon 2020's MSCA-ITN-2019 Innovative Training Networks program “Materials for Neuromorphic Circuits” (MANIC) under the grant agreement no. 861153.
M.F.S. acknowledges funding from the HORIZON-MSCA-2024-PF RECOMPUTE (101203197). M.F.S., H.R.J.C. and B.N. acknowledge funding from the HORIZON-CL4-2023 grant CONCEPT (101135946).

\section{Author contributions}
These authors contributed equally: E.vdV. and M.F.S. E.vdV. and M.F.S. performed the epitaxial thin-film synthesis with support from J.T.E., S.F., and D.vdV. E.vdV. performed the large-scale reciprocal space mapping and analyzed the structure of the films. Thin-film diffraction was performed by M.F.S. and E.vdV. with support from E.M.K. M.F.S. performed PFM and cAFM measurements. J.vG. G. performed XPS measurements. E.vdV. performed STEM acquisition and image analysis with support from M.A. and supervision from B.J.K. H.R.J.C. performed the Hall measurements. The project was conceived and coordinated jointly by E.vdV., M.F.S., and B.N. All authors participated in the analysis, discussion, and interpretation of the results. E.vdV., M.F.S., and B.N. wrote the manuscript with inputs from all authors.

\section{Conflict of interest}
The authors declare no conflicts of interest.

\section{Data availability}
The data supporting this study are available in DataverseNL at \href{https://doi.org/10.34894/QDT8NH}{https://doi.org/10.34894/QDT8NH}.

\section{Keywords}
\ch{WO3}, polarization, polar domains, epitaxy, thin film

\clearpage

\begin{singlespace}
    \printbibliography

@article{sonsteby2013,
  author  = {Sønsteby, Henrik Hovde and Chernyshov, Dmitry and Getz, Michael and Nilsen, Ola and Fjellvåg, Helmer},
  title   = {On the application of a single-crystal κ-diffractometer and a CCD area detector for studies of thin films},
  journal = {Journal of Synchrotron Radiation},
  year    = {2013},
  volume  = {20},
  number  = {4},
  pages   = {644-647},
  month   = {7}
}

@article{salje1975,
  author  = {Salje, E. and Viswanathan, K.},
  title   = {Physical Properties and Phase Transitions in \ch{WO3}},
  journal = {Acta Crystallogr. A},
  volume  = {A 31},
  number  = {May },
  pages   = {356-359},
  doi     = {Doi 10.1107/S0567739475000745},
  year    = {1975}
}

@article{woodward1995,
  author  = {Woodward, P. M. and Sleight, A. W. and Vogt, T.},
  title   = {Structure Refinement of Triclinic Tungsten Trioxide},
  journal = {J. Phys. Chem. Solids},
  volume  = {56},
  number  = {10},
  pages   = {1305-1315},
  doi     = {10.1016/0022-3697(95)00063-1},
  year    = {1995}
}

@article{lazic2016,
  title    = {Phase contrast STEM for thin samples: Integrated differential phase contrast},
  journal  = {Ultramicroscopy},
  volume   = {160},
  pages    = {265-280},
  year     = {2016},
  issn     = {0304-3991},
  doi      = {10.1016/j.ultramic.2015.10.011},
  author   = {Ivan Lazić and Eric G.T. Bosch and Sorin Lazar}
}

@article{aird1998a,
  author  = {Aird, A. and Domeneghetti, M. C. and Mazzi, F. and Tazzoli, V. and Salje, E. K. H.},
  title   = {Sheet superconductivity in \ch{WO_{3-x}}: crystal structure of the tetragonal matrix},
  journal = {J. Phys. Condens. Mat.},
  volume  = {10},
  number  = {33},
  pages   = {L569-L574},
  doi     = {10.1088/0953-8984/10/33/002},
  year    = {1998}
}

@article{aird1998b,
  author  = {Aird, A. and Salje, E. K. H.},
  title   = {Sheet superconductivity in twin walls: experimental evidence of \ch{WO_{3-x}}},
  journal = {J. Phys. Condens. Mat.},
  volume  = {10},
  number  = {22},
  pages   = {L377-L380},
  doi     = {10.1088/0953-8984/10/22/003},
  year    = {1998}
}

@article{aird2000,
  author  = {Aird, A. and Salje, E. K. H.},
  title   = {Enhanced reactivity of domain walls in \ch{WO3} with sodium},
  journal = {Eur. Phys. J. B},
  volume  = {15},
  number  = {2},
  pages   = {205-210},
  year    = {2000}
}

@article{diehl1978,
  author  = {Diehl, R. and Brandt, G. and Salje, E.},
  title   = {Crystal Structure of Triclinic \ch{WO3}},
  journal = {Acta Crystallogr. B},
  volume  = {34},
  number  = {Apr},
  pages   = {1105-1111},
  doi     = {10.1107/S0567740878005014},
  year    = {1978}
}

@article{du2014,
  author  = {Du, Y. G. and Gu, M. and Varga, T. and Wang, C. M. and Bowden, M. E. and Chambers, S. A.},
  title   = {Strain Accommodation by Facile \ch{WO6} Octahedral Distortion and Tilting during \ch{WO3} Heteroepitaxy on SrTiO3(001)},
  journal = {ACS Appl. Mater. Inter.},
  volume  = {6},
  number  = {16},
  pages   = {14253-14258},
  doi     = {10.1021/am5035686},
  year    = {2014}
}

@article{eckstein2022,
  author  = {Eckstein, J. T. and Salje, E. K. H. and Howard, C. J. and Carpenter, M. A.},
  title   = {Symmetry and strain analysis of combined electronic and structural instabilities in tungsten trioxide, \ch{WO3}},
  journal = {J. Appl. Phys.},
  volume  = {131},
  number  = {21},
  pages   = {17},
  doi     = {10.1063/5.0093803},
  year    = {2022}
}

@article{eckstein2024,
  author  = {Eckstein, J. T. and Yokota, H. and Domingo, N. and Catalan, G. and Aktas, O. and Carpenter, M. A. and Salje, E. K. H.},
  title   = {Domain wall dynamics in tungsten trioxide: Evidence for polar domain walls},
  journal = {Phys. Rev. B},
  volume  = {110},
  number  = {9},
  doi     = {10.1103/PhysRevB.110.094107},
  year    = {2024}
}

@article{garg2000,
  author  = {Garg, A. and Leake, J. A. and Barber, Z. H.},
  title   = {Epitaxial growth of \ch{WO3} films on \ch{SrTiO3} and sapphire},
  journal = {J. Phys. D: Appl. Phys.},
  volume  = {33},
  number  = {9},
  pages   = {1048-1053},
  doi     = {10.1088/0022-3727/33/9/303},
  year    = {2000}
}

@article{howard2002,
  author  = {Howard, C. J. and Luca, V. and Knight, K. S.},
  title   = {High-temperature phase transitions in tungsten trioxide - the last word?},
  journal = {J. Phys. Condens. Mat.},
  volume  = {14},
  number  = {3},
  pages   = {377-387},
  doi     = {10.1088/0953-8984/14/3/308},
  year    = {2002}
}

@article{kalhori2016,
  author  = {Kalhori, H. and Porter, S. B. and Esmaeily, A. S. and Coey, M. and Ranjbar, M. and Salamati, H.},
  title   = {Morphology and structural studies of \ch{WO3} films deposited on \ch{SrTiO3} by pulsed laser deposition},
  journal = {Appl. Surf. Sci.},
  volume  = {390},
  pages   = {43-49},
  doi     = {10.1016/j.apsusc.2016.08.052},
  year    = {2016}
}

@article{kim2010,
  author  = {Kim, Y. and Alexe, M. and Salje, E. K. H.},
  title   = {Nanoscale properties of thin twin walls and surface layers in piezoelectric \ch{WO_{3-x}}},
  journal = {Appl. Phys. Lett.},
  volume  = {96},
  number  = {3},
  pages   = {3},
  doi     = {10.1063/1.3292587},
  year    = {2010}
}

@article{leng2015,
  author  = {Leng, X. and Pereiro, J. and Strle, J. and Bollinger, A. T. and Bozovic, I.},
  title   = {Epitaxial growth of high quality \ch{WO3} thin films},
  journal = {APL Mater.},
  volume  = {3},
  number  = {9},
  doi     = {10.1063/1.4930214},
  year    = {2015}
}

@article{li2015,
  author  = {Li, G. Q. and Varga, T. and Yan, P. F. and Wang, Z. G. and Wang, C. M. and Chambers, S. A. and Du, Y. G.},
  title   = {Crystallographic dependence of photocatalytic activity of \ch{WO3} thin films prepared by molecular beam epitaxy},
  journal = {Phys. Chem. Chem. Phys.},
  volume  = {17},
  number  = {23},
  pages   = {15119-15123},
  doi     = {10.1039/c5cp01344e},
  year    = {2015}
}

@article{locherer1999,
  author  = {Locherer, K. R. and Swainson, I. P. and Salje, E. K. H.},
  title   = {Phase transitions in tungsten trioxide at high temperatures - a new look},
  journal = {J. Phys. Condens. Mat.},
  volume  = {11},
  number  = {35},
  pages   = {6737-6756},
  doi     = {10.1088/0953-8984/11/35/312},
  year    = {1999}
}

@article{mascello2020,
  author  = {Mascello, N. and Spaldin, N. A. and Narayan, A. and Meier, Q. N.},
  title   = {Theoretical investigation of twin boundaries in \ch{WO3}: Structure, properties, and implications for superconductivity},
  journal = {Phys. Rev. Res.},
  volume  = {2},
  number  = {3},
  doi     = {10.1103/PhysRevResearch.2.033460},
  year    = {2020}
}

@article{matthias1949,
  author  = {Matthias, B. T.},
  title   = {Ferro-electric Properties of \ch{WO3}},
  journal = {Phys. Rev.},
  volume  = {76},
  number  = {3},
  pages   = {430-431},
  doi     = {10.1103/PhysRev.76.430.2},
  year    = {1949}
}

@article{seo2023,
  author  = {Seo, J. and Nahm, H. H. and Park, H. S. and Yun, S. H. and Lee, J. H. and Kim, Y. J. and Kim, Y. H. and Yang, C. H.},
  title   = {Detection of the prototype symmetry of ferroelastic \ch{WO3} domain walls by angle-resolved polarized Raman spectroscopy},
  journal = {Phys. Rev. B},
  volume  = {108},
  number  = {1},
  doi     = {10.1103/PhysRevB.108.014103},
  year    = {2023}
}

@article{sun2024,
  author  = {Sun, Z. T. and Yuan, Z. Y. and Xiao, M. and Fairclough, S. M. and Jan, A. T. and Di Martino, G. and Ducati, C. and Strkalj, N. and MacManus-Driscoll, J. L.},
  title   = {Low-Temperature Epitaxy of Perovskite \ch{WO3} Thin Films under Atmospheric Conditions},
  journal = {Small Struct.},
  volume  = {5},
  number  = {7},
  doi     = {10.1002/sstr.202400089},
  year    = {2024}
}

@article{tagtstrom1999,
  author  = {Tägtström, P. and Jansson, U.},
  title   = {Chemical vapour deposition of epitaxial \ch{WO3} films},
  journal = {Thin Solid Films},
  volume  = {352},
  number  = {1-2},
  pages   = {107-113},
  doi     = {10.1016/S0040-6090(99)00379-X},
  year    = {1999}
}

@article{woodward1997,
  author  = {Woodward, P. M. and Sleight, A. W. and Vogt, T.},
  title   = {Ferroelectric tungsten trioxide},
  journal = {J. Solid State Chem.},
  volume  = {131},
  number  = {1},
  pages   = {9-17},
  doi     = {DOI 10.1006/jssc.1997.7268},
  year    = {1997}
}

@article{yang2017,
  author  = {Yang, J. T. and Ma, C. and Ge, C. and Zhang, Q. H. and Du, J. Y. and Li, J. K. and Huang, H. Y. and He, M. and Wang, C. and Meng, S. and Gu, L. and Lu, H. B. and Yang, G. Z. and Jin, K. J.},
  title   = {Effects of line defects on the electronic and optical properties of strain-engineered \ch{WO3} thin films},
  journal = {J. Mater. Chem. C},
  volume  = {5},
  number  = {45},
  pages   = {11694-11699},
  doi     = {10.1039/c7tc03896h},
  year    = {2017}
}

@article{yun2020,
  author  = {Yun, S. and Song, K. and Chu, K. and Hwang, S. Y. and Kim, G. Y. and Seo, J. and Woo, C. S. and Choi, S. Y. and Yang, C. H.},
  title   = {Flexopiezoelectricity at ferroelastic domain walls in \ch{WO3} films},
  journal = {Nat. Commun.},
  volume  = {11},
  number  = {1},
  pages   = {4898},
  doi     = {10.1038/s41467-020-18644-w},
  year    = {2020}
}

@article{yun2015,
  author  = {Yun, S. and Woo, C. S. and Kim, G. Y. and Sharma, P. and Lee, J. H. and Chu, K. and Song, J. H. and Chung, S. Y. and Seidel, J. and Choi, S. Y. and Yang, C. H.},
  title   = {Ferroelastic twin structures in epitaxial \ch{WO3} thin films},
  journal = {Appl. Phys. Lett.},
  volume  = {107},
  number  = {25},
  doi     = {10.1063/1.4938396},
  year    = {2015}
}

@article{hamdi2016,
  author  = {Hamdi, H. and Salje, E. K. H. and Ghosez, P. and Bousquet, E.},
  title   = {First-principles reinvestigation of bulk \ch{WO3}},
  journal = {Phys. Rev. B},
  volume  = {94},
  number  = {24},
  pages   = {11},
  doi     = {10.1103/PhysRevB.94.245124},
  year    = {2016}
}

@article{catalan2008,
  author  = {Catalan, G. and Béa, H. and Fusil, S. and Bibes, M. and Paruch, P. and Barthélémy, A. and Scott, J. F.},
  title   = {Fractal dimension and size scaling of domains in thin films of multiferroic \ch{BiFeO3}},
  journal = {Phys. Rev. Lett.},
  volume  = {100},
  number  = {2},
  doi     = {10.1103/PhysRevLett.100.027602},
  year    = {2008}
}

@article{pertsev1995,
  author  = {Pertsev, N. A. and Zembilgotov, A. G.},
  title   = {Energetics and Geometry of 90-Degrees Domain-Structures in Epitaxial Ferroelectric and Ferroelastic Films},
  journal = {J. Appl. Phys.},
  volume  = {78},
  number  = {10},
  pages   = {6170-6180},
  doi     = {10.1063/1.360561},
  year    = {1995}
}

@article{pompe1993,
  author  = {Pompe, W. and Gong, X. and Suo, Z. and Speck, J. S.},
  title   = {Elastic Energy-Release Due to Domain Formation in the Strained Epitaxy of Ferroelectric and Ferroelastic Films},
  journal = {J. Appl. Phys.},
  volume  = {74},
  number  = {10},
  pages   = {6012-6019},
  doi     = {10.1063/1.355215},
  year    = {1993}
}

@misc{crameri2018,
  author       = {Crameri, F.},
  title        = {Scientific colour maps},
  doi          = {10.5281/zenodo.1243862},
  year         = {2018},
  howpublished = {Zenodo}
}

@article{catalan2007,
  author  = {Catalan, G. and Scott, J. F. and Schilling, A. and Gregg, J. M.},
  title   = {Wall thickness dependence of the scaling law for ferroic stripe domains},
  journal = {J. Phys. Condens. Mat.},
  volume  = {19},
  number  = {2},
  doi     = {10.1088/0953-8984/19/2/022201},
  year    = {2007}
}

@article{schilling2006,
  author  = {Schilling, A. and Adams, T. B. and Bowman, R. M. and Gregg, J. M. and Catalan, G. and Scott, J. F.},
  title   = {Scaling of domain periodicity with thickness measured in \ch{BaTiO3} single crystal lamellae and comparison with other ferroics},
  journal = {Phys. Rev. B},
  volume  = {74},
  number  = {2},
  doi     = {10.1103/PhysRevB.74.024115},
  year    = {2006}
}

@article{maher2024,
  author  = {Maher, O. and Bernini, R. and Harnack, N. and Gotsmann, B. and Sousa, M. and Bragaglia, V. and Karg, S.},
  title   = {Highly reproducible and CMOS-compatible \ch{VO2}-based oscillators for brain-inspired computing},
  journal = {Sci. Rep.i},
  volume  = {14},
  number  = {1},
  pages   = {11600},
  doi     = {10.1038/s41598-024-61294-x},
  year    = {2024}
}

@article{boescke2011,
  author  = {Boescke, T. S. and Müller, J. and Bräuhaus, D. and Schröder, U. and Böttger, U.},
  title   = {Ferroelectricity in hafnium oxide thin films},
  journal = {Appl. Phys. Lett.},
  volume  = {99},
  number  = {10},
  doi     = {10.1063/1.3634052},
  year    = {2011}
}

@article{deb2008,
  author  = {Deb, S. K.},
  title   = {Opportunities and challenges in science and technology of \ch{WO3} for electrochromic and related applications},
  journal = {Sol. Energ. Mat. Sol. C},
  volume  = {92},
  number  = {2},
  pages   = {245-258},
  doi     = {10.1016/j.solmat.2007.01.026},
  year    = {2008}
}

@article{dong2020,
  author  = {Dong, C. and Zhao, R. and Yao, L. and Ran, Y. and Zhang, X. and Wang, Y.},
  title   = {A review on \ch{WO3} based gas sensors: Morphology control and enhanced sensing properties},
  journal = {J. Alloy Compd.},
  volume  = {820},
  doi     = {10.1016/j.jallcom.2019.153194},
  year    = {2020}
}

@article{granqvist2014,
  author  = {Granqvist, C. G.},
  title   = {Electrochromics for smart windows: Oxide-based thin films and devices},
  journal = {Thin Solid Films},
  volume  = {564},
  pages   = {1-38},
  doi     = {10.1016/j.tsf.2014.02.002},
  year    = {2014}
}

@article{begon-lours2022,
  author  = {Bégon-Lours, L. and Halter, M. and Puglisi, F. M. and Benatti, L. and Falcone, D. F. and Popoff, Y. and Dávila Pineda, D. and Sousa, M. and Offrein, B. J.},
  title   = {Scaled, Ferroelectric Memristive Synapse for Back-End-of-Line Integration with Neuromorphic Hardware},
  journal = {Adv. Electron. Mater.},
  volume  = {8},
  number  = {6},
  doi     = {10.1002/aelm.202101395},
  year    = {2022}
}

@article{ji2016,
  author  = {Ji, Y. and Yang, Y. and Lee, S. K. and Ruan, G. and Kim, T. W. and Fei, H. and Lee, S. H. and Kim, D. Y. and Yoon, J. and Tour, J. M.},
  title   = {Flexible Nanoporous \ch{WO_{3-x}} Nonvolatile Memory Device},
  journal = {ACS Nano},
  volume  = {10},
  number  = {8},
  pages   = {7598-603},
  doi     = {10.1021/acsnano.6b02711},
  year    = {2016}
}

@article{nesterov2013,
  author  = {Nesterov, O. and Matzen, S. and Magen, C. and Vlooswijk, A. H. G. and Catalan, G. and Noheda, B.},
  title   = {Thickness scaling of ferroelastic domains in \ch{PbTiO3} films on \ch{DyScO3}},
  journal = {Appl. Phys. Lett.},
  volume  = {103},
  number  = {14},
  doi     = {10.1063/1.4823536},
  year    = {2013}
}

@software{vanderveerSTEMfit,
  author  = {van der Veer, Ewout},
  license = {MIT},
  title   = {STEMfit},
  url     = {https://github.com/evanderveer/STEMfit}
}

@article{tikhonov2022,
  author  = {Tikhonov, Y. and Maguire, J. R. and McCluskey, C. J. and McConville, J. P. V. and Kumar, A. and Lu, H. and Meier, D. and Razumnaya, A. and Gregg, J. M. and Gruverman, A. and Vinokur, V. M. and Luk'yanchuk, I.},
  title   = {Polarization Topology at the Nominally Charged Domain Walls in Uniaxial Ferroelectrics},
  journal = {Adv. Mater.},
  volume  = {34},
  number  = {45},
  pages   = {e2203028},
  doi     = {10.1002/adma.202203028},
  year    = {2022}
}

@article{necas2012,
  author      = {Nečas, David and Klapetek, Petr},
  affiliation = {CEITEC — Central European Institute of Technology, Masaryk University Kamenice 753/5, 625 00 Brno, Czech Republic},
  title       = {Gwyddion: an open-source software for {SPM} data analysis},
  journal     = {Central European Journal of Physics},
  publisher   = {Versita, co-published with Springer-Verlag GmbH},
  issn        = {1895-1082},
  pages       = {181-188},
  volume      = {10},
  issue       = {1},
  year        = {2012},
  doi         = {10.2478/s11534-011-0096-2}
}

@article{schindelin2012,
  author  = {Schindelin, J. and Arganda-Carreras, I. and Frise, E. and Kaynig, V. and Longair, M. and Pietzsch, T. and Preibisch, S. and Rueden, C. and Saalfeld, S. and Schmid, B. and Tinevez, J.-Y. and White, D. J. and Hartenstein, V. and Eliceiri, K. and Tomancak, P. and Cardona, A.},
  title   = {Fiji: an open-source platform for biological-image analysis},
  journal = {Nature Methods},
  volume  = {9},
  number  = {7},
  pages   = {676-682},
  doi     = {10.1038/nmeth.2019},
  year    = {2012}
}

@incollection{zuiderveld1994,
  title     = {Contrast Limited Adaptive Histogram Equalization},
  author    = {Zuiderveld, K.},
  editor    = {Heckbert, P. S.},
  booktitle = {Graphics Gems IV},
  pages     = {474-485},
  year      = {1994},
  chapter   = {VIII.5},
  publisher = {Academic Press Professional, Inc.}
}

@article{bezanson2017,
  title     = {Julia: A fresh approach to numerical computing},
  author    = {Bezanson, Jeff and Edelman, Alan and Karpinski, Stefan and Shah, Viral B},
  journal   = {SIAM review},
  volume    = {59},
  number    = {1},
  pages     = {65-98},
  year      = {2017},
  publisher = {SIAM},
  doi       = {10.1137/141000671}
}

@software{bates2023,
  author    = {Douglas Bates and
               Andreas Noack and
               Simon Kornblith and
               Milan Bouchet-Valat and
               Michael Krabbe Borregaard and
               Alex Arslan and
               John Myles White and
               Dave Kleinschmidt and
               Phillip Alday and
               Galen Lynch and
               Iain Dunning and
               Patrick Kofod Mogensen and
               Sam Lendle and
               Dilum Aluthge and
               Mousum Dutta and
               pdeffebach and
               José Bayoán Santiago Calderón, PhD and
               Ayush Patnaik and
               Benjamin Born and
               Bradley Setzler and
               Chris DuBois and
               Jacob Quinn and
               Ondřej Slámečka and
               Paul Bastide and
               Viral B. Shah and
               Anthony Blaom, PhD and
               Bernhard König},
  title     = {JuliaStats/GLM.jl: v1.9.0},
  month     = {8},
  year      = {2023},
  publisher = {Zenodo},
  version   = {v1.9.0},
  doi       = {10.5281/zenodo.8345558}
}

@article{krumbiegel2021,
  doi       = {10.21105/joss.03349},
  year      = {2021},
  publisher = {The Open Journal},
  volume    = {6},
  number    = {65},
  pages     = {3349},
  author    = {Simon Danisch and Julius Krumbiegel},
  title     = {{Makie.jl}: Flexible high-performance data visualization for {Julia}},
  journal   = {Journal of Open Source Software}
}

@article{hunnestad2020,
  author  = {Hunnestad, K. A. and Roede, E. D. and van Helvoort, A. T. J. and Meier, D.},
  title   = {Characterization of ferroelectric domain walls by scanning electron microscopy},
  journal = {J. Appl. Phys.},
  volume  = {128},
  number  = {19},
  doi     = {10.1063/5.0029284},
  year    = {2020}
}

@article{woodward1997a,
  author  = {Woodward, P. M.},
  title   = {Octahedral Tilting in Perovskites. I. Geometrical Considerations},
  journal = {Acta Crystallogr. B},
  year    = {1997},
  volume  = {53},
  number  = {1},
  pages   = {32-43},
  month   = {2},
  doi     = {10.1107/S0108768196010713}
}

@article{glazer1972,
  author  = {Glazer, A. M.},
  title   = {The classification of tilted octahedra in perovskites},
  journal = {Acta Crystallogr. B},
  year    = {1972},
  volume  = {28},
  number  = {11},
  pages   = {3384-3392},
  month   = {11},
  doi     = {10.1107/S0567740872007976}
}

@article{mcgilly2017,
  author  = {McGilly, Leo J. and Feigl, Ludwig and Setter, Nava},
  title   = {Dynamics of ferroelectric \ang{180} domain walls at engineered pinning centers},
  journal = {Appl. Phys. Lett.},
  volume  = {111},
  number  = {2},
  pages   = {022901},
  year    = {2017},
  month   = {7},
  doi     = {10.1063/1.4993576}
}

@article{matsumoto2010,
  author  = {Matsumoto, Takao and Okamoto, Masakuni},
  journal = {IEEE Trans. Ultrason. Ferroelectr. Freq. Control},
  title   = {Ferroelectric \ang{180} a-a nanostripe and nanoneedle domains in thin \ch{BaTiO3} films prepared with focused-ion beam},
  year    = {2010},
  volume  = {57},
  number  = {10},
  pages   = {2127-2133},
  doi     = {10.1109/TUFFC.2010.1668}
}

@article{hassani2022,
  title     = {First-principles study of lattice dynamical properties of the room-temperature \emph{P2\textsubscript{1}/n} and ground-state \emph{P2\textsubscript{1}/c} phases of \ch{WO3}},
  author    = {Hassani, Hamideh and Partoens, Bart and Bousquet, Eric and Ghosez, Philippe},
  journal   = {Phys. Rev. B},
  volume    = {105},
  number    = {1},
  pages     = {014107},
  year      = {2022},
}

@article{hassani2025,
  title   = {The anti-distortive polaron as an alternative mechanism for lattice-mediated charge trapping},
  author  = {Hassani, Hamideh and Bousquet, Eric and He, Xu and Partoens, Bart and Ghosez, Philippe},
  journal = {Nat. Commun.},
  volume  = {16},
  pages   = {1688},
  doi     = {10.1038/s41467-025-56791-0},
  year    = {2025}
}

@article{Nath2010,
  author  = {Nath, Ramesh and Hong, Seungbum and Klug, Jeffrey A. and Imre, Alexandra and Bedzyk, Michael J. and Katiyar, Ram S. and Auciello, Orlando},
  title   = {Effects of cantilever buckling on vector piezoresponse force microscopy imaging of ferroelectric domains in \ch{BiFeO3} nanostructures},
  journal = {Appl. Phys. Lett.},
  volume  = {96},
  number  = {16},
  pages   = {163101},
  year    = {2010},
  month   = {04},
  issn    = {0003-6951},
  doi     = {10.1063/1.3327831},
}

@article{Gradauskaite2023,
  title   = {Defeating depolarizing fields with artificial flux closure in ultrathin ferroelectrics},
  author  = {Gradauskaite, Elzbieta and Meier, Quintin N. and Gray, Natascha and Sarott, Martin F. and Scharsach, Tizian and Campanini, Marco and Moran, Thomas and Vogel, Alexander and Del Cid-Ledezma, Karla and Huey, Bryan D. and Rossell, Marta D. and Fiebig, Manfred and Trassin, Morgan},
  journal = {Nat. Mater.},
  volume  = {22},
  pages   = {1492-1498},
  doi     = {10.1038/s41563-023-01674-2},
  year    = {2023}
}

@article{Nataf2020,
  title   = {Domain-wall engineering and topological defects in ferroelectric and ferroelastic materials},
  author  = {Nataf, Guillaume F. and Guennou, M. and Gregg, J.M. and Meier, D. and Hlinka, J. and Salje, E.K.H. and Kreisel, J.},
  journal = {Nat. Rev. Phys.},
  volume  = {2},
  pages   = {634-648},
  doi     = {10.1038/s42254-020-0235-z},
  year    = {2020}
}

@article{vanRijn2024,
  title   = {Domains with Varying Conductance in Tensile Strained \ch{SrMnO3} Thin Films Using Out-of-Plane Electric Fields},
  author  = {van Rijn, Job J. L. and Bhaduri, Ishitro and Ahmadi, Majid and Noheda, Beatriz and Kooi, Bart J. and Banerjee, Tamalika},
  journal = {Adv. Funct. Mater.},
  volume  = {34},
  pages   = {2404150},
  doi     = {10.1002/adfm.202404150},
  year    = {2024}
}

@article{Shih1997,
  title   = {Secondary electron emission studies},
  author  = {A. Shih and J. Yater and C. Hor and R. Abrams},
  journal = {Appl. Surf. Sci.},
  volume  = {111},
  pages   = {251-258},
  doi     = {10.1016/S0169-4332(96)00729-5},
  year    = {1997}
}

@article{Natanzon2020,
  author   = {Natanzon, Yuriy and Azulay, Amram and Amouyal, Yaron},
  title    = {Evaluation of Polaron Transport in Solids from First-principles},
  journal  = {Isr. J. Chem.},
  volume   = {60},
  number   = {8-9},
  pages    = {768-786},
  doi      = {10.1002/ijch.201900101},
  year     = {2020}
}

@book{lindemuth2020hall,
  title     = {Hall Effect Measurement Handbook: A Fundamental Tool for Semiconductor Material Characterization},
  author    = {Lindemuth, J. and Dodrill, B.},
  isbn      = {9781734707809},
  year      = {2020},
  publisher = {Lake Shore Cryotronics}
}

@software{CoxGithub,
  author = {Cox, Horatio R. J.},
  title  = {Electronic Properties of Thin Films},
  url    = {https://github.com/hc3413/Electronic_properties_of_thin_films}
}

@article{Catalan2012,
  title     = {Domain wall nanoelectronics},
  author    = {Catalan, G. and Seidel, J. and Ramesh, R. and Scott, J. F.},
  journal   = {Rev. Mod. Phys.},
  volume    = {84},
  issue     = {1},
  pages     = {119-156},
  numpages  = {0},
  year      = {2012},
  month     = {2},
  doi       = {10.1103/RevModPhys.84.119},
}

@article{Meier2012,
  author   = {Meier, D. and Seidel, J. and Cano, A. and Delaney, K. and Kumagai, Y. and Mostovoy, M. and Spaldin, N. A. and Ramesh, R. and Fiebig, M.},
  title    = {Anisotropic conductance at improper ferroelectric domain walls},
  journal  = {Nat. Mater.},
  year     = {2012},
  month    = {4},
  day      = {01},
  volume   = {11},
  number   = {4},
  pages    = {284-288},
  issn     = {1476-4660},
  doi      = {10.1038/nmat3249},
}

@article{Seidel2009,
  author   = {Seidel, J. and Martin, L. W. and He, Q. and Zhan, Q. and Chu, Y.-H. and Rother, A. and Hawkridge, M. E. and Maksymovych, P. and Yu, P. and Gajek, M. and Balke, N. and Kalinin, S. V. and Gemming, S. and Wang, F. and Catalan, G. and Scott, J. F. and Spaldin, N. A. and Orenstein, J. and Ramesh, R.},
  title    = {Conduction at domain walls in oxide multiferroics},
  journal  = {Nat. Mater.},
  year     = {2009},
  month    = {3},
  day      = {01},
  volume   = {8},
  number   = {3},
  pages    = {229-234},
  issn     = {1476-4660},
  doi      = {10.1038/nmat2373},
}
\end{singlespace}

\clearpage
\section*{Supplementary material}
\clearpage

\setcounter{figure}{0}
\renewcommand{\thefigure}{S\arabic{figure}} 

\begin{figure}
    \centering
    \includegraphics[width=\textwidth]{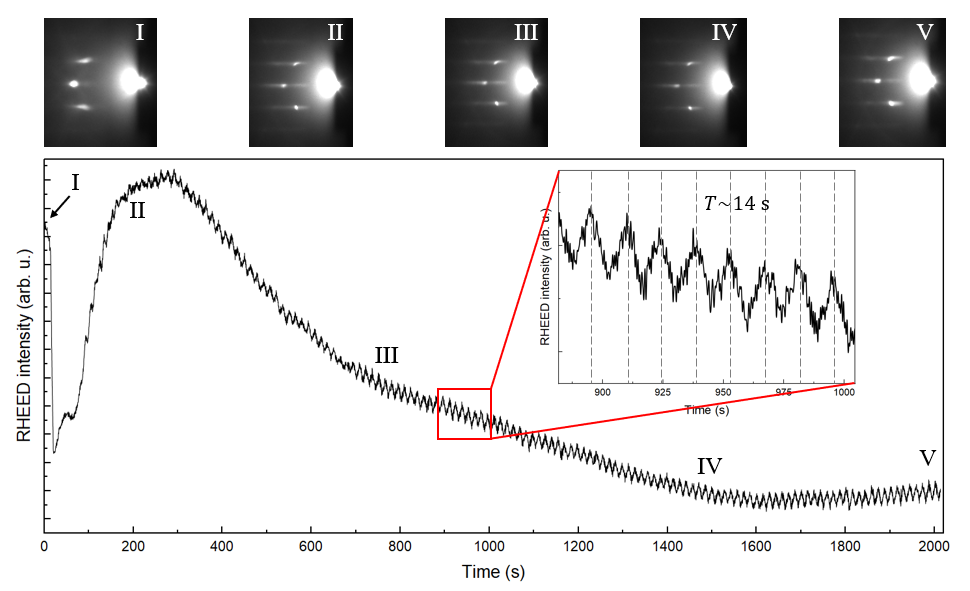}
    \caption{Example RHEED pattern of the PLD growth of the \ch{WO3} films.}
    \label{fig:suppl:rheed}
\end{figure}

\begin{figure}
    \centering
    \includegraphics[width=\textwidth]{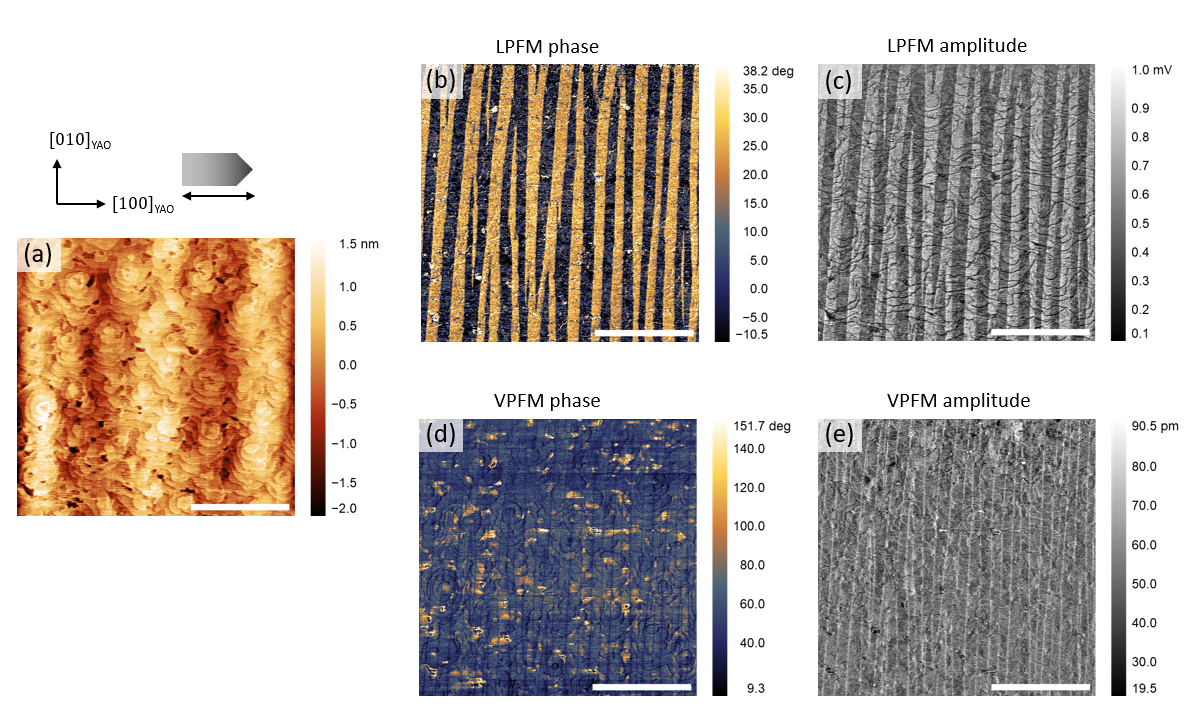}
    \caption{Lateral and vertical PFM at a sample rotation of \ang{0}.}
    \label{fig:suppl:pfm0deg}
\end{figure}
\begin{figure}
    \centering
    \includegraphics[width=\textwidth]{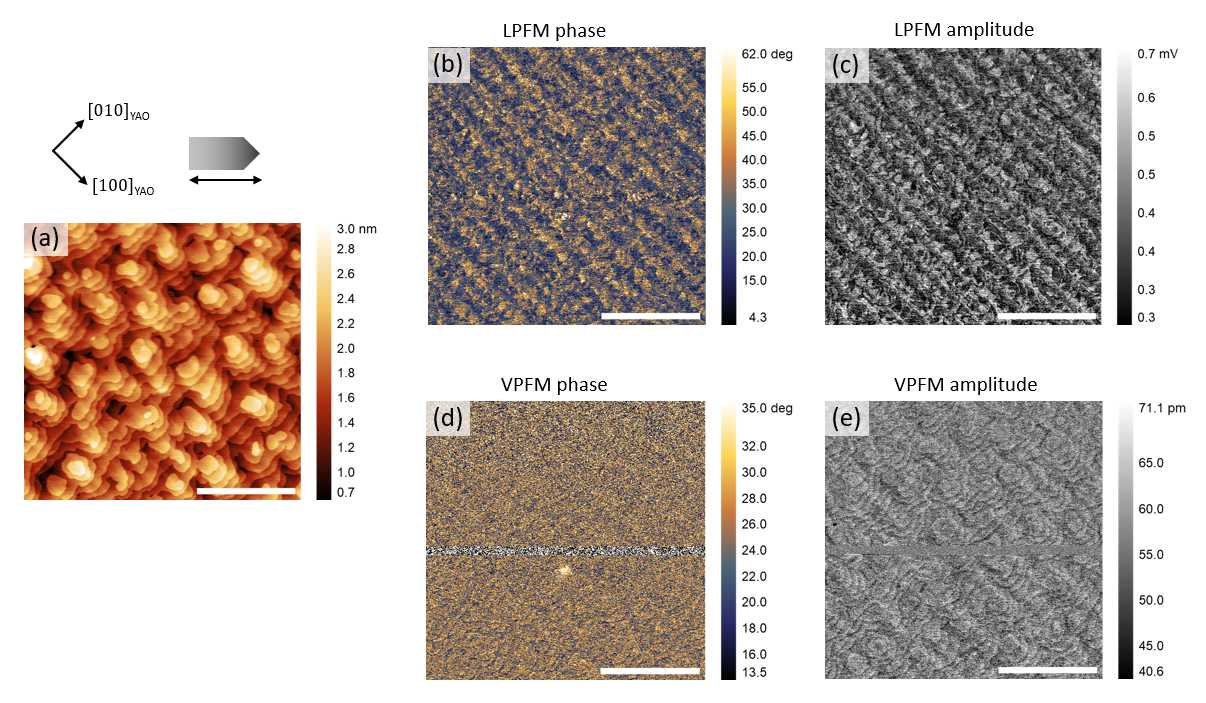}
    \caption{Lateral and vertical PFM at a sample rotation of \ang{45}.}
    \label{fig:suppl:pfm45deg}
\end{figure}
\begin{figure}
    \centering
    \includegraphics[width=\textwidth]{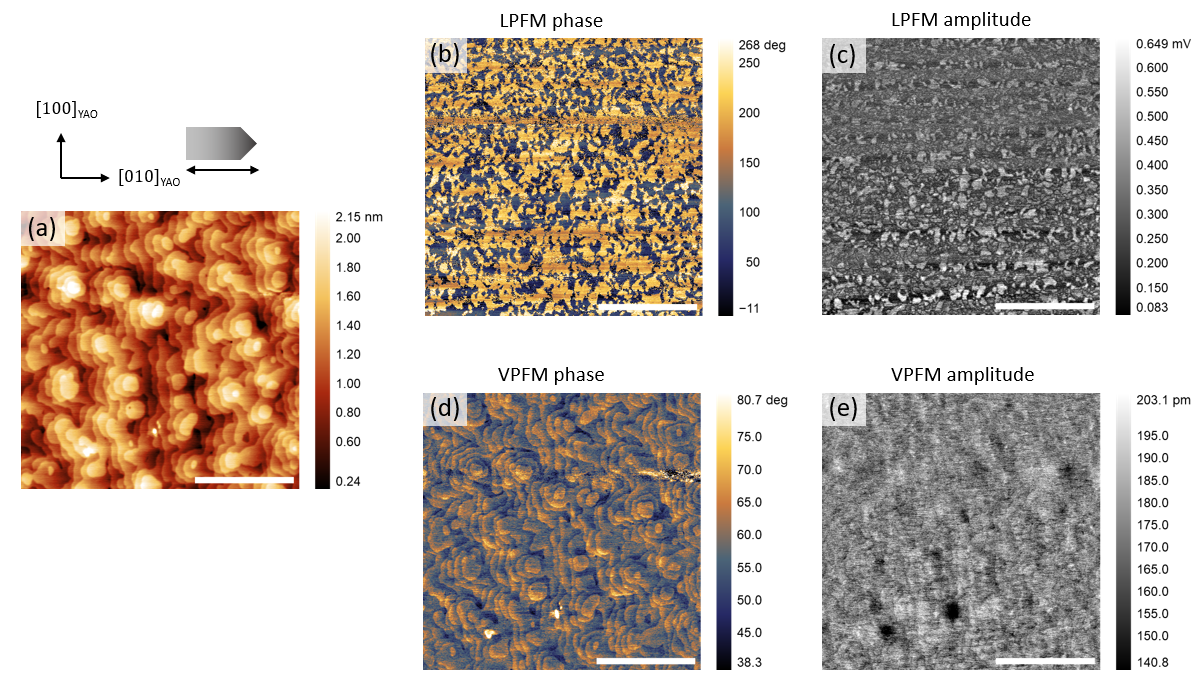}
    \caption{Lateral and vertical PFM at a sample rotation of \ang{90}.}
    \label{fig:suppl:pfm90deg}
\end{figure}

\begin{figure}
    \centering
    \includegraphics[width=\textwidth]{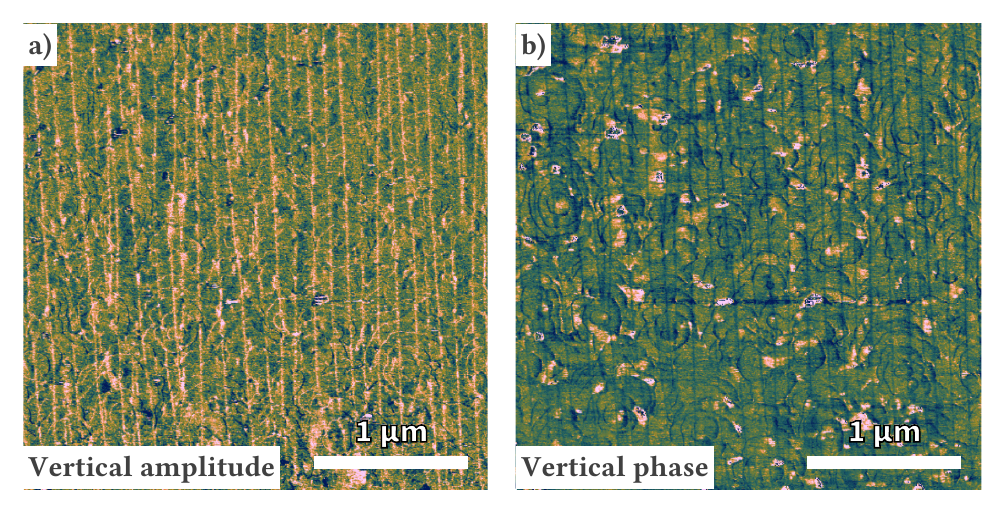}
    \caption{Vertical PFM a) amplitude and b) phase corresponding to the data shown in Fig.\ \ref{fig:pfm}.}
    \label{fig:suppl:vertpfm}
\end{figure}

\begin{figure}
    \centering
    \includegraphics[width=\textwidth]{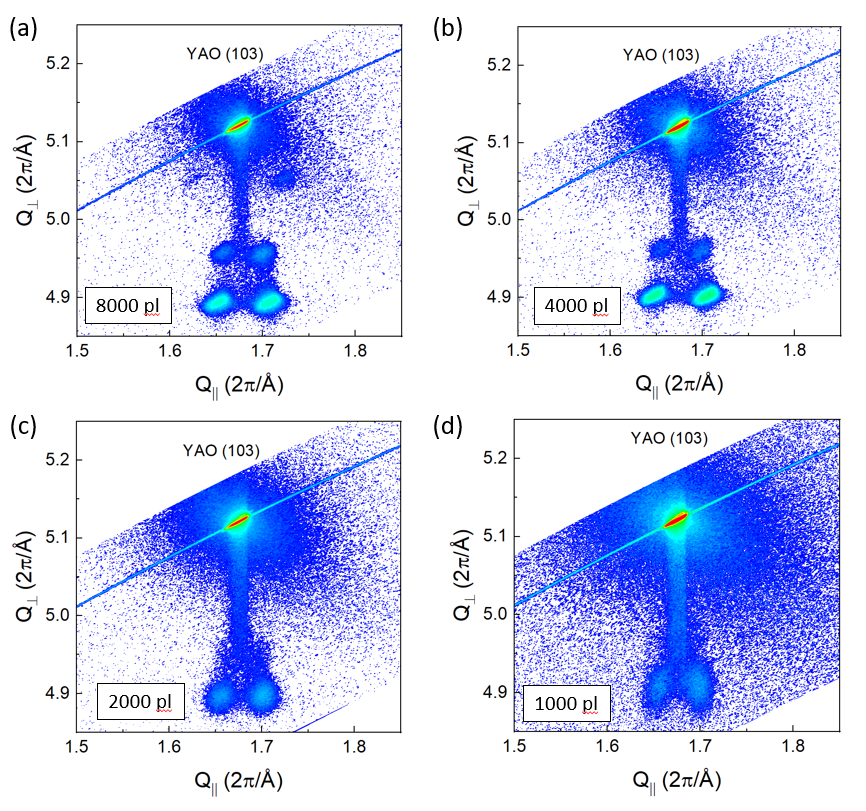}
    \caption{Reciprocal space maps close to the pseudocubic (103) reflections of the YAO substrate of \ch{WO3} films with four different thicknesses.}
    \label{fig:suppl:rsm_thicknesses}
\end{figure}
\begin{figure}
    \centering
    \includegraphics[width=\textwidth]{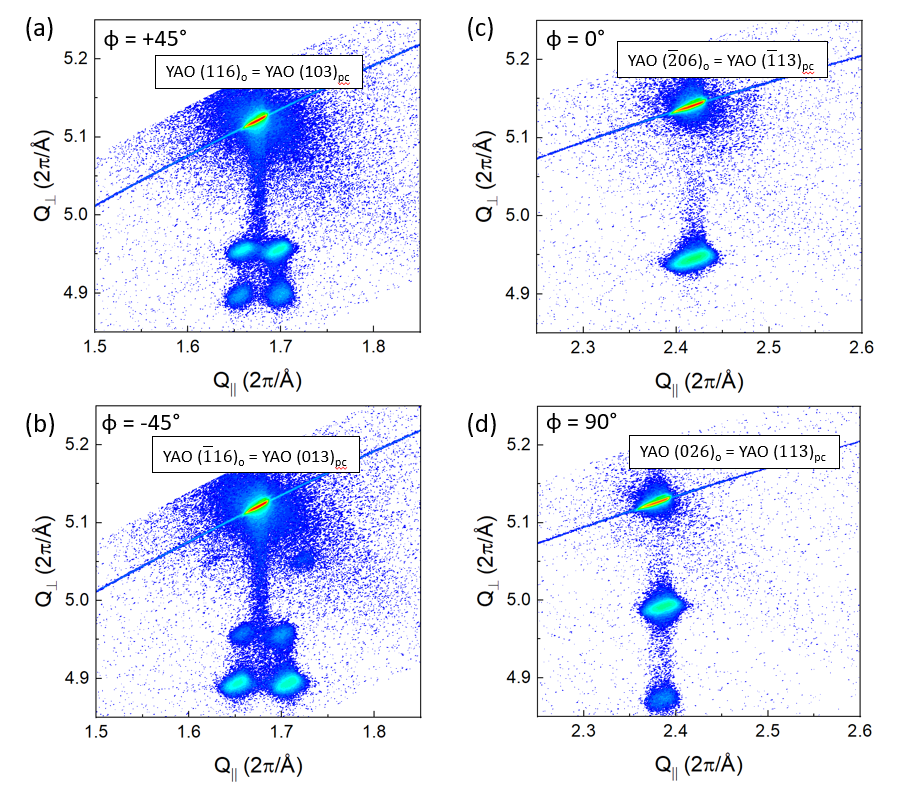}
    \caption{Reciprocal space maps close to the pseudocubic (103) reflections of the YAO substrate in four different azimuthal orientations.}
    \label{fig:suppl:rsm_orientations}
\end{figure}

\begin{figure}
    \centering
    \includegraphics[width=\textwidth]{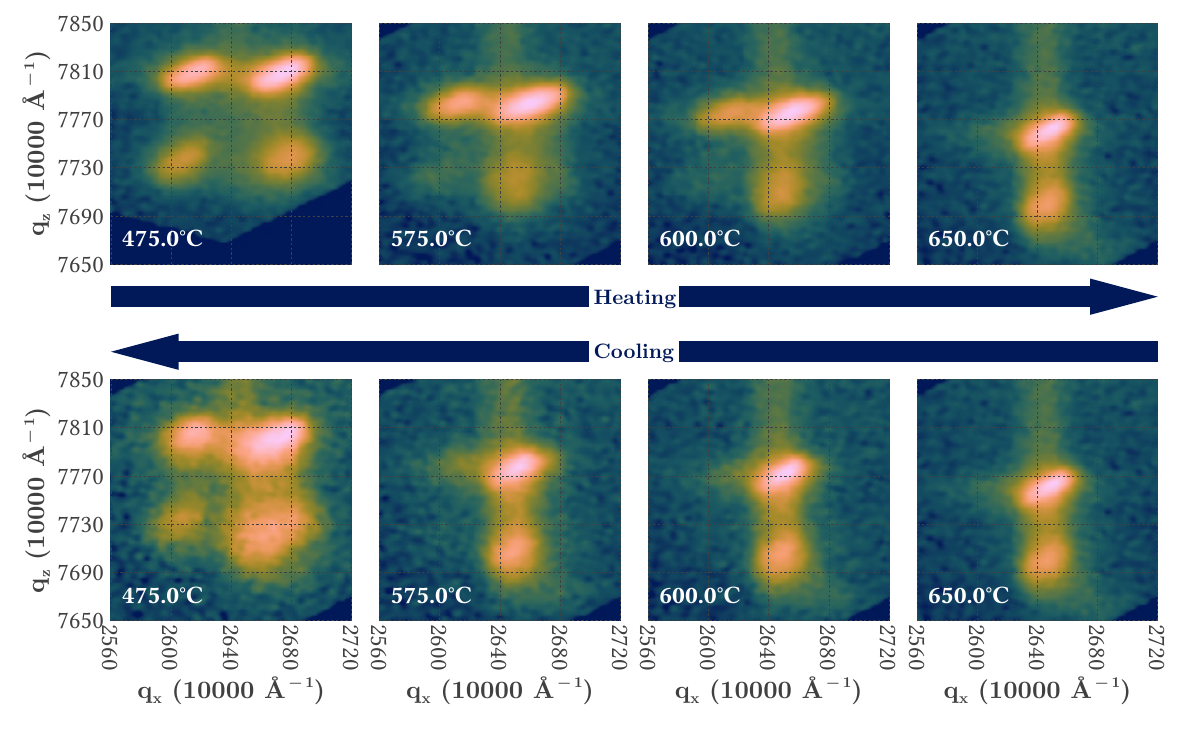}
    \caption{Reciprocal space map of the pseudo-cubic {103} reflections of the \ch{WO3} film at different temperatures through the phase transition.}
    \label{fig:suppl:rsms}
\end{figure}

\begin{figure}
    \centering
    \includegraphics[width=\textwidth]{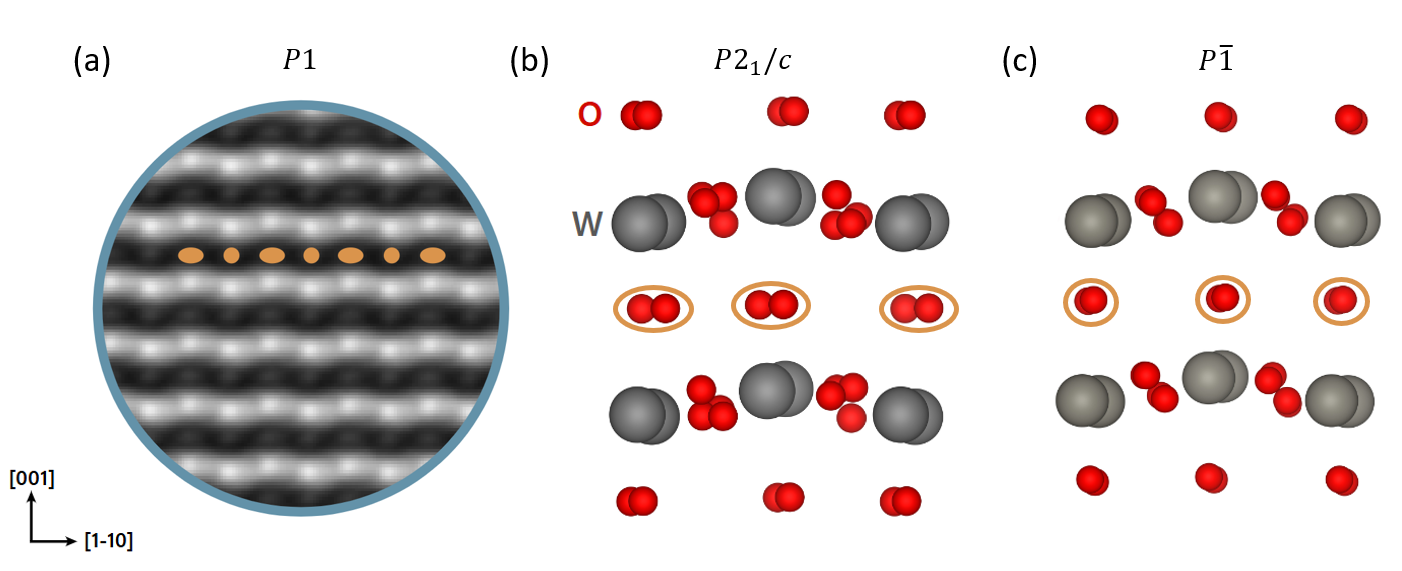}
    \caption{Unit cells of the (a) \emph{P2\textsubscript{1}/n} and (b) \emph{P-1} structures found in bulk \ch{WO3} in the (110) zone. Note that all atomic columns of oxygen in the \vao sublayers are identical, unlike in our films. }
    \label{fig:suppl:bulkstructure}
\end{figure}

\begin{figure}
    \centering
    \includegraphics[width=0.6\textwidth]{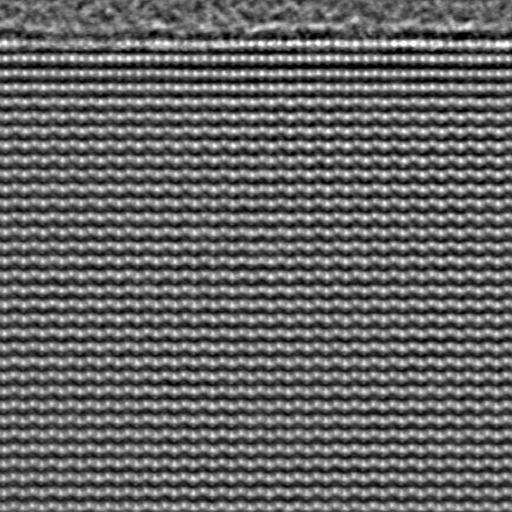}
    \caption{iDPC-STEM image close to the film surface showing a change of the structure similar to that seen close to the domain walls.}
    \label{fig:suppl:stem_surface}
\end{figure}

\begin{figure}
    \centering
    \includegraphics[width=\textwidth]{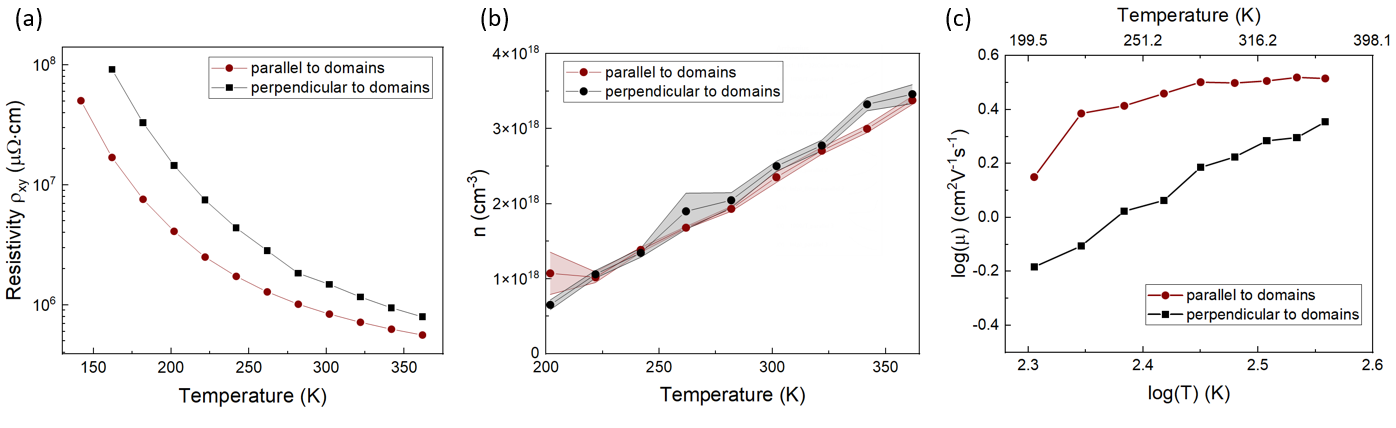}
    \caption{a) Resistivity, b) carrier concentration and c) mobility extracted from Hall bar measurements.}
    \label{fig:suppl:hall}
\end{figure}

\begin{figure}
    \centering
    \includegraphics{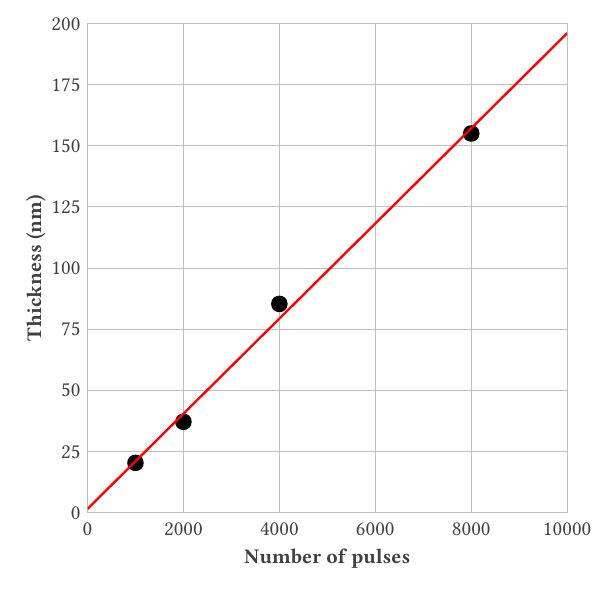}
    \caption{Thickness of the \ch{WO3} films grown by pulsed layer deposition on (001)\ch{YAlO3} substrates, as a function of the number of laser pulses used. The data points were measured from the spacing of Laue oscillations in figure \ref{fig:structural}(c). The thickness values given in the main text were obtained from the linear fit shown in red. }
    \label{fig:suppl:thickness}
\end{figure}

\begin{figure}
    \centering
    \includegraphics[width=\textwidth]{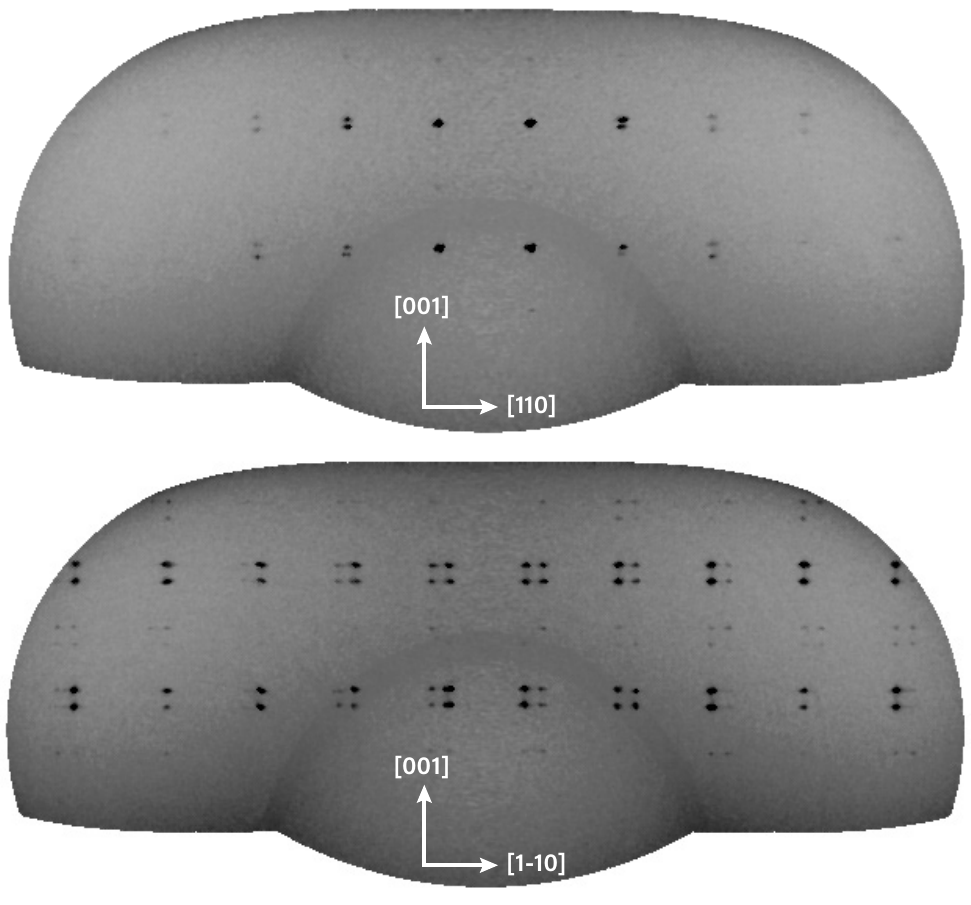}
    \caption{Reconstructed reciprocal lattice planes as seen along the [1\textoverline{1}0] (top panel) and [110] (bottom panel). These planes correspond to $\mathrm{n=4}$ in figure \ref{fig:structural}(a,b).}
    \label{fig:suppl:latticeplanes}
\end{figure}

\begin{figure}
    \centering
    \includegraphics{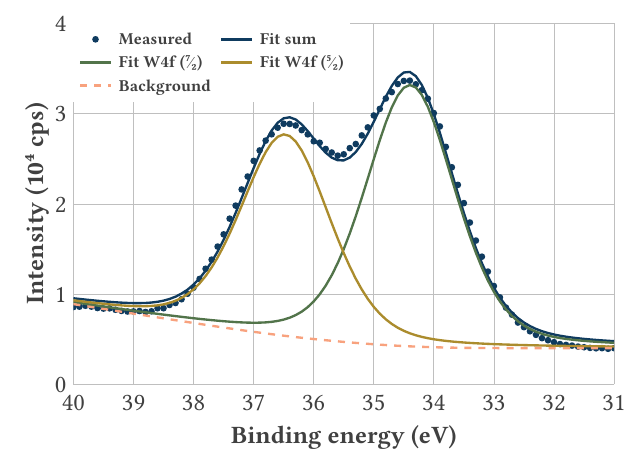}
    \caption{X-ray photoelectron spectrum of the W4f peak showing a single valence state for tungsten. The lack of W\textsuperscript{5+} implies that the film is near-stoichiometric.}
    \label{fig:suppl:xps}
\end{figure}

\end{document}